\documentclass[12pt,latexsym]{amsart}
\usepackage{url}

\begin{document}

\title{Symplectic Quantum Mechanics and Chern-Simons Gauge Theory I}

\author{Lisa C. Jeffrey}
\address{Department of Mathematics \\
University of Toronto \\ Toronto, Ontario \\ Canada}
\email{jeffrey@math.toronto.edu}
\urladdr{\url{http://www.math.toronto.edu/~jeffrey}} 

\maketitle

\begin{abstract}
In this article we describe the relation between 
the Chern-Simons gauge theory partition function and
the partition function defined using the 
symplectic action functional as the Lagrangian.
We show that the partition functions obtained using these two Lagrangians
agree, and we identify the semiclassical formula for the partition
function defined using the symplectic action functional.
\end{abstract}

%


\def\baselinestretch{1.5}

\setlength{\parskip}{0.2\baselineskip}


\newcommand{\bC}{{\bf C}}
\newcommand{\bZ}{{\bf Z}}
\newcommand{\brpr}{\bar{\partial}}
\newcommand{\pr}{\partial}
\newcommand{\al}{\alpha}
\newcommand{\ga}{\gamma}
\newcommand{\Ga}{\Gamma}
\newcommand{\eps}{\epsilon}
\newcommand{\la}{\lambda}
\newcommand{\La}{\Lambda}
\newcommand{\om}{\omega}
\newcommand{\Om}{\Omega}
\newcommand{\Si}{\Sigma}
\newcommand{\si}{\sigma}
\newcommand{\nb}{\nabla}
\newcommand{\tlnab}{\tilde{\nabla}}

\newcommand{\lrar}{\longrightarrow}
\newcommand{\Tr}{\,{\rm Tr}\,} 
\newcommand{\End}{\,{\rm End}\,}
\newcommand{\Hom}{\,{\rm Hom}\,}
\newcommand{\Ker}{ \,{\rm Ker} \,}
 
\newcommand{\bla}{\phantom{bbbbb}} 
\newcommand{\onebl}{\phantom{a} }
\newcommand{\eqdef}{\;\: {\stackrel{ {\rm def} }{=} } \;\:}
\newcommand{\sign}{\: {\rm sign}\: }
\newcommand{\sgn}{ \:{\rm sgn}\:}                                      
\newcommand{\half}{ {\frac{1}{2} } }
\newcommand{\vol}{ \,{\rm vol}\, }

%

\newcommand{\beq}{\begin{equation}}
\newcommand{\eeq}{\end{equation}}
\newcommand{\beqst}{\begin{equation*}}
\newcommand{\eeqst}{\end{equation*}}
\newcommand{\barr}{\begin{array}}
\newcommand{\earr}{\end{array}}
\newcommand{\beqar}{\begin{eqnarray}}
\newcommand{\eeqar}{\end{eqnarray}}
\newtheorem{theorem}{Theorem}[section]

\newtheorem{lemma}[theorem]{Lemma}
\newtheorem{prop}[theorem]{Proposition}
\newtheorem{definition}[theorem]{Definition}
\newtheorem{remit}[theorem]{Remark}

\newcommand{\matr}[4]{\left \lbrack \begin{array}{cc} #1 & #2 \\
     #3 & #4 \end{array} \right \rbrack} 
\newcommand{\colvec}[2]{\left \lbrack \begin{array}{cc} #1  \\
     #2  \end{array} \right \rbrack}

\newenvironment{rem}{\begin{remit}\rm}{\end{remit}}

-----------------------

\newcommand{\aff}{{\mbox{$\mathbb {A}$}}}
\newcommand{\RR}{{\mbox{$\mathbb {R}$}}}
\newcommand{\CC}{{\mbox{$\mathbb {C}$}}}
\newcommand{\ZZ}{{\mbox{$\mathbb {Z}$}}}

\newcommand{\cala}{{\mbox{$\mathcal A$}}}
\newcommand{\calb}{{\mbox{$\mathcal B$}}}
\newcommand{\calc}{{\mbox{$\mathcal C$}}}
\newcommand{\cald}{{\mbox{$\mathcal D$}}}
\newcommand{\cale}{{\mbox{$\mathcal E$}}}
\newcommand{\calf}{{\mbox{$\mathcal F$}}}
\newcommand{\calg}{{\mbox{$\mathcal G$}}}
\newcommand{\calh}{{\mbox{$\mathcal H$}}}
\newcommand{\cali}{{\mbox{$\mathcal I$}}}
\newcommand{\calj}{{\mbox{$\mathcal J$}}}
\newcommand{\calk}{{\mbox{$\mathcal K$}}}
\newcommand{\call}{{\mbox{$\mathcal L$}}}
\newcommand{\calm}{{\mbox{$\mathcal M$}}}
\newcommand{\caln}{{\mbox{$\mathcal N$}}}
\newcommand{\calo}{{\mbox{$\mathcal O$}}}
\newcommand{\calp}{{\mbox{$\mathcal P$}}}
\newcommand{\calq}{{\mbox{$\mathcal Q$}}}
\newcommand{\calr}{{\mbox{$\mathcal R$}}}
\newcommand{\cals}{{\mbox{$\mathcal S$}}}
\newcommand{\calt}{{\mbox{$\mathcal T$}}}
\newcommand{\calu}{{\mbox{$\mathcal U$}}}
\newcommand{\calv}{{\mbox{$\mathcal V$}}}
\newcommand{\calw}{{\mbox{$\mathcal W$}}}
\newcommand{\calx}{{\mbox{$\mathcal X$}}}
\newcommand{\caly}{{\mbox{$\mathcal Y$}}}
\newcommand{\calz}{{\mbox{$\mathcal Z$}}}

\newcommand{\qqed}{\hfill \mbox{$\Box$}\medskip\newline}

\newcommand{\U}{U}
\newcommand{\tlbe}{{\tilde{\beta}}}
\newcommand{\ftil}{\tilde{f}}
\newcommand{\atil}{{\tilde{A}}}
\newcommand{\ad}{{\rm ad}}
\newcommand{\cs}{CS}
\newcommand{\lamax}{\Lambda^{\rm max}}
\newcommand{\lieg}{{\bf g}}
\newcommand{\liet}{{\bf t}}
\newcommand{\ddtau}{\frac{\partial}{\partial \tau}}
\newcommand{\nbt}{\nabla^{(t)} }
\newcommand{\bunt}{\widetilde{TN}}
\newcommand{\buntpr}{\widetilde{T^*N}}
\newcommand{\ddt}{\frac{\partial}{\partial t}}

\newcommand{\barpit}{{\bar{\Pi_t}}}

\newcommand{\tder}{\frac{d}{d t}}
\newcommand{\barpi}{{\bar{\Pi}}}
\newcommand{\tlu}{{\tilde{u} }}
\newcommand{\dbar}{\bar{\partial} }
\newcommand{\vardelta}{\delta}
\newcommand{\buntdopr}{\tilde{T''N }}

\newcommand{\tld}{\widetilde{D}}
\newcommand{\sigone}{\Sigma_1}
\newcommand{\sigthr}{\Sigma_3}
\newcommand{\sigtwo}{\Sigma_2}
\newcommand{\SF}{SF}
\renewcommand{\Tr}{{\rm Tr}}
\newcommand{\lineb}{{\mathcal{L}}}
\newcommand{\tf}{\tilde{f}}
\newcommand{\tlineb}{\tilde{\lineb}}
\newcommand{\hatx}{E}
\newcommand{\A}{A}
\newcommand{\abs}[1]{||#1||}

\newcommand{\rootl}{\Lambda^{\rm R}}
\newcommand{\weightl}{\Lambda^{\rm W}}
\newcommand{\N}{V}
\newcommand{\splin}{SL(2, \ZZ)}
\renewcommand{\frak}{\mathfrak}
\newcommand{\factr}{ (p-w-w^{-1}) }
\newcommand{\absc}{|c|}
\newcommand{\vpm}{|d + a \pm 2|} 
\newcommand{\posrts}{|\Delta_+|}

\newcommand{\htil}{{\tilde{H}}}
\newcommand{\hsig}{\calh(\Sigma)}
\newcommand{\lf}{{\call}_f}
\newcommand{\dph}{\dot{\gamma}}
\newcommand{\tilu}{{\tilde{u}}}
\newcommand{\dps}{\dot{\psi}}
\newcommand{\liesp}{{\mathfrak{sp}}}
\newcommand{\txomf}{T_{x_0} \Omega_f}
\newcommand{\txom}{{T_{x_0} \Omega^0}}
\newcommand{\ddu}{\frac{\partial}{\partial u}}
\newcommand{\tlxt}{\widetilde{X}_t}
\newcommand{\sympl}{Sp(2n, \RR)}
\newcommand{\rea}{\RR}
\newcommand{\zbar}{{\bar{z}}}
\newcommand{\zbari}{\zbar_i}
\newcommand{\zbarj}{\zbar_j}
\newcommand{\keradd} {{\rm Ker (ad)}}
\newcommand{\imadd}{{\rm Im(ad)}}
\newcommand{\coe}{\lambda}
\newcommand{\xhat}{E}
\newcommand{\zj}{Z_j}
\newcommand{\zk}{Z_k}
\newcommand{\jm}{j}
\newcommand{\tpr}{t'}
\newcommand{\dfn}{:=}
\newcommand{\bps}{\bar{\psi}}
\newcommand{\expad}{{\rm exp(ad) }}
\newcommand{\jb}{\bar{j}}
\newcommand{\sfd}

\newcommand{\be}{\beta}

\pagestyle{plain}

\tableofcontents
\section{Introduction} \label{intro}

\subsection{Chern-Simons gauge theory} \label{intro:csgt}
This article  deals partly  with 
{\it Chern-Simons gauge theory}, a topological field 
theory in $2 + 1 $ dimensions.  
As shown by Witten, this theory 
leads  to invariants  of 
three-manifolds $M$.
The data necessary to define such invariants are
a compact Lie group $G$ (which we shall assume { simple}
and {simply connected}, and much of the time shall take to be
$G = SU(2)$), together with
a parameter $k$, the {\it level} or {\em coupling constant}.
The standard technique in physics for treating 
path integrals  is to evaluate them as power series
in $\frac{1}{k}$, using stationary phase 
approximation around the critical points
of the Lagrangian. 

To define the invariant $Z(M,k)$ of a three-manifold $M$, 
one uses the space $\cala$ of connections $A$ on a principal 
$G$-bundle $P$ over $M$. (As $G$ is simply connected, 
$P$ must be trivial.)  The space of connections 
$\cala$ is the space of fields  in the theory; the Lagrangian is the 
{\it Chern-Simons functional}
\beq \cs(A) 
= \frac{1}{8 \pi^2} \int_M \Tr (A \, dA + \frac{2}{3} A^3 ).\eeq
We shall refer to $Z(M,k)$ as the {\it partition function}
or the {\it Chern-Simons-Witten} (CSW) {\it invariant}.
 The Chern-Simons functional was first used in  
the physics literature in  \cite{Deser}. See also  \cite{Dunne}.

\subsection{Stationary phase approximation}

We assume we have defined the partition function
\beq \label{pathint}
Z(k) = \int \cald \ga \: e^{i k L(\ga) } . \eeq
Here,
$\cald \ga$ denotes 
the path integral measure at a point $\ga$ in the space 
of fields $\Om$ with  Lagrangian  $L: \Om \to \RR$.
The coupling constant
$k$ is usually a real number, but it will be an integer
in all the examples we shall consider, since actually 
our Lagrangians will take values in $\RR/2 \pi \ZZ$.   

One computes    the leading  order contribution 
to $Z(k) $ by {\it stationary phase approximation}. 
This is by analogy with integrals of this form 
over $\RR^n$ (see  \cite{GS}). The basic idea is 
that as $k \to  \infty $, the integrand will oscillate 
wildly and average to zero, except near the 
critical points $\ga_j$ of $L$. One then 
expands $L$ to second order at $\ga_j$. 
In other words,  we consider the tangent space 
$T_{\ga_j} \Om$, which is assumed equipped with 
an inner product $( \, \cdot , \cdot \, )$. 
Further, we assume there is a map 
$$ \exp : T_{\ga_j} \Om \to \Om . $$
Thus we may define an operator $D_j \in \End(T_{\ga_j} \Om) $ 
(for $\xi \in T_{\ga_j} \Om$) by 
$$L(\exp \, \xi) = L(\ga_j) + (\xi, D_j \xi )  +    
{\rm terms \onebl of \onebl higher \onebl order}. $$

The leading order contribution to the path integral is now 
defined by analogy with the finite-dimensional case. If 
$D$ is an invertible operator on $\RR^n$, we 
analytically continue the  
integral $\int d^n x \, e^{- (x, Dx)} $ for positive
definite operators $D$ to obtain 
\beq \label{eq:stph}
\int d^n x \, e^{i (x, Dx) } = \pi^{n/2} \, |\det D |^{-1/2}
\:
e^{i \frac{\pi }{4} \sign D }.  \eeq
The operators 
$D$ arising in quantum field theory  are typically 
elliptic differential operators; one may define a determinant 
by zeta function regularization, or by the 
method outlined in \S \ref{s:det}. 

The signature of an operator also has an 
analogue in infinite dimensions: this is the 
eta invariant $\eta(D)$. 
Suppose $D$ is a  
non-positive self-adjoint elliptic operator. 
If $\lambda$ are
the eigenvalues, one may define
$$\eta(s) = \lim_{s \to 0} \sum_\lambda {\rm sign} \lambda \: 
|\lambda|^{-s} $$
for large ${\rm Re}(s)$. This extends to a meromorphic function
which is in fact holomorphic at $s=0$.
We define the eta invariant  
$\eta(D)$ to be   the value at $0$ of this extended function. 
The eta invariant  is a measure of the asymmetry of the
eigenvalues about 0; if they occur in pairs $\pm \lambda$,  $\eta(D)$
vanishes.  Unlike the signature of a finite dimensional operator, $\eta(D)$
is  in general not an integer.

Thus to leading order in $\frac{1}{k}$, the path integral 
(\ref{pathint}) is:
\beq \label{stphgen}
Z(k) \sim \sum_j |\det D_j |^{-1/2} \, e^{ik L(\ga_j)} \, 
e^{i \frac{\pi}{4} \eta(D_j) } . \eeq

\subsection{The stationary phase approximation to Chern-Simons}
The stationary phase approximation to 
 the Chern-Simons path 
integral was derived by Witten \cite{W:J}. 
The Chern-Simons path integral is defined by 
\beq \label{ch2:zpath}
Z(M,k)  = \int \cald A \: e^{2 \pi i k \cs(A) }. 
\eeq 
The critical points of the 
action $\cs(A)$ are the {\it flat connections} $A_\al$. 

The operator $D_\al $ that appears is   the Atiyah-Patodi-Singer
operator $D_\al = \ast d_{A_\al} + d_{A_\al} \ast $ 
acting on the space $\Om^{ {\rm odd} } (M, \ad \, P)$ of 
odd dimensional forms with coefficients in the flat 
bundle $P$. The factor $|\det D_\al |^{-1/2}$  is the square root
of the {\it Reidemeister torsion}  $\tau_{A_\al}$. 
Thus to leading order in $\frac{1}{k}$,
the contribution to the 
Chern-Simons partition 
function 
from isolated irreducible flat connections 
is as follows. This formula gives the leading term 
provided all the flat connections are isolated and at least 
one of them is irreducible. The product connection is 
of course reducible, but its contribution is of 
higher order in $\frac{1}{k}$. 
\beq \label{csstph}
Z(M,k) \sim \sum_\al e^{\frac{i \pi}{4} \eta (D_\al) } \;
\tau_{A_\al}^{1/2} \; e^{2 \pi i k \cs(A_\al) } . \eeq
This formula requires certain corrections, and the precise stationary
phase formula is (\ref{sp}) below.

\subsection{The path integral prediction for {\it Z(k)} } 
A recurring theme in this article is the rigorous 
verification of the path integral 
prediction (\ref{stphgen}) for    $Z(k)$. 
(\ref{stphgen}) suggests that as $k \to \infty$, 
the functional dependence of $Z(k)$ on $k$ will take  
a particular form. This form is superficially quite different
from the form in which the parameter $k$ enters the expressions
for $Z(M,k)$ obtained from quantization of 
the Chern-Simons theory. 
 These expressions appear naturally as polynomials 
in an $N$-th root of unity, where $N$ is a multiple of $r$ and 
\beq r = k + h; \eeq
here, the dual Coxeter number
$h$ is a natural integer constant associated to the 
Lie group $G$.

\subsection{The symplectic action functional}

A further recurring theme in this article is the analogy 
between the Chern-Simons functional and 
another functional, the {\it symplectic action 
functional} $S$. In the variant we shall consider, 
the symplectic action functional is defined 
given a symplectic manifold ($N, \om$) and a 
symplectic diffeomorphism $f: N \to N$. It is 
defined on the space $\Om_f$ of paths $\ga$  in $N$
such that $\ga(t + 1) = f \ga(t)$. The critical 
points of $S$ are the constant paths 
at the fixed points $x_j$ of $f$. The action $S$ is then defined 
as the integral of $\om$ over a strip between $\ga $ and 
some reference  point $x_0$   (one of the fixed 
points). 

The {\it gradient flow} of $S$ on $\Om_f$ shares 
many of the properties of the gradient flow  of 
$\cs$ on $\cala$. Both can be used to define 
a \lq\lq Floer homology" 
\cite{Flo}, in which the critical 
points form a  basis for a chain complex and 
the gradient flow trajectories define the 
boundary maps. Given the remarkable results  
obtained by treating the Chern-Simons functional 
as the action in a quantum field theory and defining
the partition function as a path  integral 
over $\cala$, it is natural to ask what 
one may obtain by defining a 
path integral over $\Om_f$ with the symplectic
action functional as the action. This is the 
content of our \S \ref{chap1}. We 
treat a field theory in $0+1$ dimensions based 
on the symplectic action functional: 
we refer to  this as {\it symplectic 
quantum mechanics} (SQM).

An important special case occurs when 
the symplectic manifold $N$ is the 
moduli space $\calm$ of 
flat connections on a $G$ bundle 
over a surface $\Si$, and $f : \calm \to \calm$ 
is induced from a diffeomorphism $\beta: \Si \to \Si$. 
Then the symplectic action functional $S$ is 
closely related to the Chern-Simons functional on the 
mapping torus 3-manifold $\Si_\beta$ formed 
by gluing $\Si \times I$ via $\beta$. The paper \cite{DS} 
shows the Floer homologies of the two functionals are 
essentially the same. One objective 
of the present article 
 is to establish the relation between 
the Chern-Simons path integral (\ref{ch2:zpath}) and 
our SQM path integral, at least in the large $k$ limit.  

\subsection{Framings and regularization} \label{introfreg}

A final theme in this article is a technique for 
using the index theorem to obtain a 
metric independent expression 
from the stationary phase expression (\ref{stphgen}), 
which appears to have some metric dependence. 
In the Chern-Simons case, for instance, one 
must introduce a metric $g$ on 
the three-manifold $M$ in order to define the 
operators $D$. The determinant of $D$ gives 
rise to Reidemeister torsion, 
which is independent of $g$; however, the 
eta invariants $\eta (D)$ depend 
nontrivially on $g$.  This behaviour occurs also 
in symplectic quantum mechanics, where the analogue 
of a metric choice is a choice of complex 
structure on the tangent spaces at the fixed points 
$T_{x_j} N$.

The regularization  technique we describe was 
introduced by Witten \cite{W:J}; here, we adapt it to 
treat the symplectic action functional.
Throughout, we are treating an index problem on a  bundle $E$ over a 
manifold $Y \times I$, which is equivalent 
to the spectral flow of a 
one parameter family of operators on $Y$.
For the Chern-Simons case, $Y$ is our three-manifold $M$; 
for the symplectic action functional, $Y$ is $S^1$.

The technique consists of two steps. First, we choose 
a reference critical point $\ga_0$ and 
consider the difference of eta invariants 
$\eta(\ga_j) -  \eta(\ga_0)$. Via the APS index theorem 
for manifolds with boundary, this may 
be expressed as follows. This formula is valid provided
the corresponding operators have no kernels.
\beq \label{indbdy}
-\frac {1}{2} ( \eta(\ga_j) -  \eta(\ga_0) )
= {\rm SF} - \int_{Y \times I} \Om,  \eeq
where ${\rm SF} $ is the spectral flow of a family of 
operators and $\Om $ is some characteristic form.
In general,  $\Om$  is a polynomial in the curvatures
of  appropriate connections on $E$ and  on  the 
tangent bundle.

For a general action $L$ on a manifold $Y$, the procedure is as follows. One 
chooses a reference point $\ga_0$ among the 
critical points of this action, and writes
$$Z = e^{\frac{i \pi}{4} \eta(\ga_0) } 
\sum_{\ga\in {\rm Crit}(L)}  \frac{ 
e^{i \pi  (\eta(\ga) - \eta(\ga_0) )/4 } e^{i k L(\ga)} }
{\vert \det D_\ga\vert^{1/2} }. $$
Here $D_\ga $ is the operator corresponding to the 
quadratic part of the Lagrangian at $\ga$. 
Now one uses the index theorem  to 
convert the eta invariant factor into 

\beq \label{sqmreg}
Z = e^{i \pi \eta(\ga_0)/4 } \sum_{\ga}  \frac{ 
e^{i \pi  \{\int_{Y \times I} \al - \dim \Ker D_\ga /2 - \SF(\ga, \ga_0) 
\}/2  } \: e^{i k L(\ga)}    }
{\vert \det D_\ga \vert^{1/2} }.  \eeq
Here, $\SF(\ga, \ga_0)$ is 
the spectral flow of a
family of operators $D_\tau    $ along  a path  
(parametrized by $\tau$) connecting
$D_\ga$ and $D_{\ga_0}$.

In the Chern-Simons case, the term $\int \Om$ leads
to a shift in  the coefficient multiplying 
$\cs (A_\al)$ in (\ref{csstph}), from $k$ to $k + h$. In the 
case of the symplectic action 
functional, it leads (under  certain hypotheses)
to a  standard result from geometric 
quantization: namely, that the line bundle $\call^k$ 
(whose curvature form is integrated to get 
the symplectic action functional) should 
be replaced by $\call^k \otimes K^{1/2} $,
where $K$ is the canonical bundle. 
We apply this to the symplectic 
manifold $\calm(\Si)$ to rederive the 
shift $k \to k + h $ from this point of view. 

The second step in the regularization 
procedure is to replace $\eta (\ga_0)$ 
(which  now  appears only in an overall factor 
multiplying the partition 
function) by some other 
quantity which is independent of the metric 
but depends on a choice of trivialization  $\rho$ 
of an appropriate    bundle  over $Y$.  For the Chern-Simons 
functional, this means a trivialization of the 
tangent bundle $TM$ (a framing of $M$).

Suppose our operator $D$ on $Y$ is extended  to 
$D_X$ over  a manifold 
$X$ with boundary $Y$. 
Given such a boundary trivialization,  one may define a 
relative characteristic number corresponding
to the 
integral $\int_X \Om(\rho) $ evaluated using the curvature 
of a connection restricting on the boundary to the 
product connection determined by the 
trivialization $\rho$. We may also consider 
$\int_{Y \times I} \Om (\nb, \rho), $ where we evaluate
$\Om $ using a connection interpolating 
between the product connection and the connection 
$\nb $ corresponding to the operator $D$. 
Then we may define 
$$\delta (D, \rho) =  \int_{Y \times I } \Om (\nb, \rho)
- \frac{\eta(D)}{2} ; $$
by the index theorem, this is equal to 
$$\delta(D, \rho) = {\rm Ind} (D_X)  - \int_X \Om (\rho). $$  
It is thus natural to  obtain a metric 
independent expression by replacing  $ - \frac{\eta (D)}{2}$ 
by $\delta(D, \rho)$. In general, what we have 
specified here might change by an integer under metric choices.
The defect $\delta$ used in the Chern-Simons case is obtained 
using the {\it signature} formula (\cite{APS}, 
I, Th.\ 4.14) rather than the index formula, (I, Th. 3.10), 
and is metric independent: see \cite{A2} and \cite{APS}.

\subsection{Summary of results}
Our work        treats the Chern-Simons partition function 
for mapping tori  $\Si_\be$ of surfaces $\Si$. 
In \S   \ref{relscs}-\ref{regul} we treat
a general surface $\Si$. Our aim is to  demonstrate the equality
of 
the Chern-Simons 
partition function 
(\ref{ch2:zpath}) for  $\Si_\be$
with the 
partition function  for a different field theory (symplectic
quantum mechanics, or 
SQM) for the corresponding 
moduli space $\calm$,
at least in the large $k$ limit. 
We do this by comparing the quantities involved 
in the stationary phase expansions. Most of 
these quantities have been studied before 
\cite{DS}, \cite{RSW}.

The expressions we seek to identify are the Chern-Simons 
stationary phase result (\ref{sp}) and the 
corresponding SQM stationary phase result 
from  (\ref{finregsqm}). We have identified the 
individual terms contributing to the two partition 
functions; however, we have not obtained a complete 
identification of the overall expressions, because the 
phases of these terms 
contain powers of $i$ that are determined by the spectral
flow of certain operators, and it is not clear that these 
coincide in general. See, however, the final 
paragraph of \S \ref{specflow}.

\noindent{\bf Notation:}
\begin{enumerate}

\item Recall that for a surface $\Sigma$ and a diffeomorphism
$\be: \Si \to \Si$, the {\it mapping torus} 
$\Si_\be$ is defined as 
$$\Si_\be = \Si \times [0,1] / \; (x, 0) \sim (\be (x), 1). $$

\item The surface $\Si$ is assumed equipped with a 
(topologically trivial) principal $G$ bundle
$P$, and we assume a lift $\tlbe$ of $\be$ to $P$
has been chosen. One may then define 
a mapping torus bundle $P_\be \to \Si_\be$, as the
mapping torus of $P$ under $\tlbe$.
For reference, we specify a flat connection
$A_0$ on $\Si$ such that $\tlbe^\ast A_0 = A_0$.
For instance, one may do this by picking a trivialization
of $P$, and taking $\tlbe$ to be the corresponding
trivial lift of $\be$ and $A_0$ to be the 
product connection.

\item We denote by $\calm$ or $\calm(\Si)$ the moduli space of 
flat connections on $P_\Si$. The map induced by 
$\tlbe$ on $\calm$ is denoted $f_\be$ or usually 
just $f$. A lift to the prequantum line 
bundle $\call $ over $\calm $ is denoted 
$\ftil$.  We denote by $\calm(\Si_\be)$  the moduli 
space of flat connections on the mapping torus 
$\Si_\be.$ \qqed
 \end{enumerate}

The remainder of this article is organized as follows.
\S \ref{chap1} is concerned with the  path integral 
using the symplectic action functional: the main 
result is the calculation of the stationary 
phase approximation to the path integral 
(\S \ref{s:det} and \ref{s:eta}). 
If the diffeomorphism $f$ preserves a compatible 
complex structure, then the stationary phase
approximation is exact (Proposition \ref{stexact}) and the partition
function is the expansion at fixed points from the 
holomorphic Lefschetz fixed point 
formula \ref{hollef}.

 \S  \ref{relscs}
treats connections on mapping tori and identifies the Chern-Simons 
path integral formally with the SQM path integral. 
\S  \ref{llim} identifies the Chern-Simons invariants and 
Reidemeister torsion of flat connections on 
mapping tori. In \S  \ref{regul} we discuss the 
regularization 
procedure outlined in \S \ref{introfreg} . 
We discuss the  
procedure for SQM in general and the particular 
case of SQM for moduli spaces of flat connections 
on surfaces. 
For comparison,  in \S  \ref{frameta} we provide
a brief exposition of the results of 
Witten \cite{W:J} on the Chern-Simons case.

The main results in this article are

\begin{enumerate}
\item The SQM partition function 
is equal to the Chern-Simons partition function of the mapping torus
(Proposition \ref{chsqfo} and \S \ref{llim})
\item proof of the explicit formula for the contribution of 
fixed points to the Chern-Simons partition function of a 
mapping torus (Proposition 5.6, \cite{lenspa} where it was stated
without proof -- the proof is given in Proposition \ref{proplenspa} below )
\item semiclassical formula for the SQM partition function, in the 
large $k$ limit  -- see \S \ref{chap1}
\end{enumerate}

{\em Remark:} Much of  material in this article
derives from the author's D. Phil. thesis \cite{J:thesis}.
Some 
other results from this thesis have already been published in \cite{lenspa}.

\section{Symplectic Quantum Mechanics} \label{chap1}
\subsection{Introduction}
This section describes a  field theory in $0+1$ dimensions.
 The space of fields is $\Om_f$, the  space of paths in a 
symplectic manifold $(N,\om)$ with endpoint condition 
imposed by a symplectic diffeomorphism $f$. The action is 
the symplectic action functional defined by Floer \cite{Flo}. 
The properties of this functional are very similar to 
those of the Chern-Simons functional; one motivation 
for this investigation was thus to shed light 
on the Chern-Simons path integral.

Our most concrete results are about the stationary phase approximation. 
In particular we provide an argument (Proposition\ \ref{stexact}) 
showing that if $f$ preserves a compatible holomorphic structure 
in addition to the symplectic structure, then the stationary 
phase approximation is exact. The proof uses Darboux's theorem to 
rewrite the action in a form that is precisely quadratic. 
The values  obtained from the  stationary phase approximation 
yield the expansion at fixed points of the 
holomorphic Lefschetz formula (\ref{hollef}).

We calculate  the stationary phase approximation  at the 
critical points  of the action, which are the constant 
paths at the fixed points $x_j$ of $f$. 
The absolute values of the determinants
that appear are independent of the complex structure, 
while the phases (eta invariants) depend on a 
choice of complex structure at the 
$x_j$. Similar behaviour is observed in the Chern-Simons
path integral \cite{W:J}. 

In order to proceed beyond the leading order in perturbation 
theory, it is necessary to transform the action into 
a function on a linear space. One method involves 
a one parameter family of metrics and 
the associated exponential maps. We opted instead to assume 
our symplectic diffeomorphisms came from the integral 
of a (time dependent) Hamiltonian flow, and to use this 
Hamiltonian flow to transform our 
action  (\ref{s1def}) into an action (\ref{eq:hamact}) 
on the space of {\it closed}
loops in $N$.

We find that there are nontrivial contributions to the partition 
function beyond the leading order in perturbation theory. This is
not surprising, since such corrections appear also in 
the Chern-Simons path integral \cite{BN}. What is more 
disappointing is that our results depend on the  choice of   a 
Darboux coordinate system near the fixed points. A possible 
reason for this is that we implicitly make use of the Euclidean 
metric induced by the Darboux coordinate system. This 
is a somewhat unnatural metric choice, since the 
symplectic manifolds in question are usually not 
globally flat (with the obvious exception of tori).

This section  is organized as follows. 
The definition of  the symplectic action 
functional and its Hessian
are given in \S \ref{s:def},  where we also 
 reformulate the path integral in terms 
of Hamiltonian flows. 
The absolute value of the determinant is 
computed in \S \ref{s:det}, while 
\S \ref{s:eta} computes the eta invariant. 

 \S \ref{s:cartan} provides background, in terms of a 
decomposition of  $Sp(2n,\RR)$ into its Cartan subgroups.

The stationary phase calculation is used 
for the special case when $N$ is the 
moduli space $\calm$ of flat connections on 
a surface.  We show that the SQM stationary phase 
calculation  agrees with the 
 Chern-Simons stationary phase  calculation  for 
mapping tori of surfaces.

A basic reference on 
background  material in symplectic 
geometry is 
 \cite{McDS}.

\subsection{ The symplectic action functional} \label{s:def}
\subsubsection{Several definitions of the action }
The symplectic action functional is a functional defined on 
paths in symplectic manifolds satisfying certain endpoint
conditions. It has a number of related definitions. The basic 
data in the situation we shall consider are a symplectic
manifold $(N,\om)$ equipped with a symplectic
diffeomorphism $f$.

The usual definition is the definition used in \cite{DS} and  \cite{Flo}.
  We define a space of paths in $N$ by 
the periodicity conditions
\beq \label{omf}
\Om_f = \{ \ga: \RR \to N \mid \; \; \ga(t + 1) = f \ga(t) \}. \eeq
The tangent space to $\Om_f$ at $\ga$ is 
\beq \label{tanomf}
T_\ga \Om_f = \{ \xi(t) \in T_{\ga(t)} N \onebl \onebl
(t \in \RR) \mid \; \xi(t+1) = f_\ast \xi(t) \}. 
\eeq 
We pick a reference point $x_0 \in N$ which is a fixed point
of $f$. The constant path at $x_0$ will be denoted 
$\ga_{x_0}$, or sometimes simply $x_0$. 
 We define the action $S_1: \Om_f \to \RR/2 \pi \ZZ $ by 
\beq \label{s1def}
S_1(\ga) = - \int_{u(I \times I)} \om, \eeq
where 
$u: I \times \RR \to N$ satisfies 
\beq \label{ucond}
u(\tau, t + 1) = f u(\tau, t), \onebl \onebl \onebl \eeq
\beq u (\tau = 0, t) = x_0, \onebl \onebl u(\tau=1, t) = \ga(t). \eeq
It is easy to check that $S_1$ is well defined 
independently of the choice of the strip $u$. 

The following is obvious from this 
description:
\begin{lemma} \label{s1diff}
The differential of $S_1$ is 
$$(dS_1)_\ga(\xi) = - \int_0^1 \om(\dot{\ga}, \xi) \, dt. $$
Thus critical points of $S_1$ correspond to fixed points 
$x_0$ 
of $f$. $\square$ 
\end{lemma}

\noindent{\bf Remark:} There are several other 
 endpoint conditions
on paths  which permit the definition of a symplectic 
action functional: see \cite{Flo}.

The following is another way to view the   symplectic
action functional. We now assume: 
\begin{enumerate}
\item $N$ is equipped with a choice of prequantum 
line bundle $(\call, \nb)$. 
\item The diffeomorphism $f$ lifts to a unitary
action $\ftil$ on $\call$ preserving 
$\nb$. (If a lift $\ftil$ exists, it is easily seen 
to be unique up to a constant $U(1)$ factor, 
provided $N$ is connected. The existence of $\ftil$ 
is guaranteed, for instance, by the hypothesis that  $N$ is simply
connected.)
\end{enumerate} 
Consider the  parallel
transport map $\Ga_\ga$  in $\call$ along $\ga 
\subset N $ from $\ga(0)$ to $\ga(1)$.
 The
action $S$ is then defined by 
\beq \label{sactdef} e^{-iS(\ga)} \, = (\Gamma_\ga \,v , \, \tf v), \eeq
where $v$ is any element of unit length in $\call_{\ga(0)}$ .

This 
definition may be reformulated as follows. 
The {\it mapping torus}
$N_f$ is the space (fibered over $S^1$) formed by  
identifying $(0,a)$ with $(1, f(a) )$ in $I \times N$. One may form a bundle $\call_f$
over $N_f$ by taking $I \times\call$ and identifying $(0,l)$
with $(1, \tf (l) )$. Because $\tf^\ast \Theta = \Theta$ (viewing 
$\call$ as a principal bundle), 
$\lf$ has a connection with curvature 
$-i \om$ along the fibres. (This 
statement makes sense since $f^\ast \om = \om$.)
The space of paths $\Om_f$  identifies with the space of sections 
of the fibration $N_f$ over $S^1$.

\begin{rem}
Observe that we have specified more structure here than simply
the diffeomorphism type of the mapping torus of $f$. As we have 
constructed it, the fibration $N_f \to S^1 $ is equipped 
with a vector field $X$ that lifts the standard 
vector field on $S^1 $. The flow along the integral curves 
of this vector field gives the monodromy map $f$. 
We are also specifying a 
lift of $X$ to the bundle $\call_f \to N_f $ (viewed as
a principal $U(1)$ bundle) such that $\iota_X \Theta = 
L_X \Theta = 0. $ 
\end{rem} 
\begin{rem}
See \cite{A} and \cite{AH} for another treatment of Chern-Simons invariants of mapping tori.
\end{rem}

We may now check:

 \begin{lemma}
The action $\exp(iS(\ga))$ is the same as the conjugate of the  
holonomy of $\lf$ around $\ga$ viewed as a path in $N_f$.  
\end{lemma}
\noindent{\bf Proof:} Suppose $s$ is a section of $\call$ in a 
neighbourhood of the path $\ga$ in $N$. Let $\theta_s$ be the form given by 
$\nabla s = i \theta_s s$. The formula for parallel transport is as
follows. If 
$$\nabla_{\dot{\ga} } \, (e^{ig} s ) = 0  
= i \left (dg(\dot{\ga})\,  +  \theta_s(\dph)\right ) (e^{ig} s)$$
then
$$g(\ga(t) ) = - \int \theta_s(\dph) \, dt,$$ i.e., the parallel
transport $\Gamma_\ga s(0) = \exp(-i\int \theta_s (\dph) \, dt ) s(1)$. 

We now restrict to paths in some neighbourhood $U$ of a fixed point
$x_0$ of $f$, on which a section $s$  of $\call $ is defined and 
also a function $h: \, U \to \RR$ such that 
$$\tf \circ s \circ f^{-1} (x) = e^{ih(x)} \, s(x).$$
Thus the action is :
\beq \label{eq:act}
e^{-iS(\ga)} = 
\exp \left \lbrace -i  \int  \left ( \theta_s (\dph)  + 
h(\ga(1)\, \right ) \, dt 
\right \rbrace.                                                       \eeq
We compare this with the holonomy around $\lf$. 
One may define a section 
$\tilde{s} : \, I \times N \to \call$  of $\lf $ as follows:
$$\tilde{s} (t, a) \: = \: e^{ith(a)} s(a). $$
Then the pullback $\nabla(\tilde{s} ) \,= \, i \theta_{\tilde{s} }  \,
\tilde{s}  $ is given by 
$$\theta_{\tilde{s} } = d(th) + \theta_s$$
so that as in the above calculation the holonomy around $\ga$ is 
$$\exp \left \lbrace -i \int \theta_{\tilde{s} } (\dph) \, dt 
\right \rbrace \,= \, 
\exp \left \lbrace  -i \int \theta_s (\dph) -i h (\ga(1) ) \right \rbrace , $$
which is exactly our previous expression
(\ref{eq:act}).  \hfill $\square$

It is natural to choose a family of complex structures $J_t$ 
on $N$ such that 
\beq \label{jtimed}
f^\ast J_{t+1}(x) = df(x)^{-1} \, J_{t+1}\,\bigl ( \,f(x) \, \bigr  )
 \, df(x) \: 
= J_t(x). 
\eeq 
This has the effect that if $ \xi = [\xi(t)]  \in T_\ga \Om_f$, then 
$[J_t \xi (t)] \in T_\ga \Om_f$ as well. 

One may 
use the corresponding metrics
$g_t$  to  define an exponential 
map $T_\ga \Om_f \to \Om_f.$   
We have 
$$f^\ast g_{t+1} = g_t, $$
so defining  
the corresponding exponential maps $\exp_t$, 
we have 
\beq \label{expend}\exp_{t+1}\circ df = f \circ \exp_t . \eeq
We may  thus define $\exp : T_\ga \Om_f \to \Om_f$ by 
(for $\xi \in T_\ga \Om_f$) 
\beq \label{expomf}
(\exp \xi)(t) = \exp_t \xi(t). \eeq

\begin{lemma} \label{agre}
The two definitions $S_1$ (from (\ref{s1def})) and $S$ (from 
(\ref{sactdef})) of the symplectic action
functional agree.
\end{lemma}
\noindent{\bf Proof:} 
Let $\ga_0$, $\ga_1$ $\in \Om_f$, and take a strip $u(t, \tau)$ in 
$N$ satisfying  (\ref{ucond}),  such that $u(t, 0) = \ga_0(t), $
and $u(t, 1) = \ga_1(t). $
We now view $\ga_0$, $\ga_1$ as 
sections of $N_f$,  and define a strip $\tilu$ in $N_f$ 
with oriented boundary  $ -\ga_1 \cup \ga_0$ in the obvious way, by 
\beq \label{tiludef}
\tilu(t, \tau) = \bigl  ( t, u(t, \tau) \,\bigr  ). \eeq  
Then the difference of holonomies is 
$$
 - i \int \theta_s \,(\dot{\ga_1} )  + 
i \int \theta_s \,(\dot{\ga_0} )
 = i \int_{\tilu} \omega, $$
so 
\beq 
S(\ga_1) - S(\ga_0) = - \int_{\tilu} \om = - \int_u \om. \eeq
This agrees with the definition (\ref{s1def}) of
$S_1.$ \hfill $\square$

The Hessian of $S$ at a critical point $\ga= x_0$ is the 
symmetric quadratic form on 
$T_{\ga} \Om_f$  given by
$$ (\nb^2 S)_\ga (\xi, \eta) = - \int_0^1 \om
\left (\dot{\xi}(t), \eta(t) \right ).   \bla \square $$ 
Thus near a critical point, to second order in $\xi \in 
T_{x_0} \Om_f$, we have 
$$S(\exp \xi) = S(x_0) +  \frac{1}{2} (\nb^2 S)_{\ga_{x_0}} (\xi, \xi). $$

\begin{prop} \label{stexact}
If $f$ preserves a metric on $N$ compatible 
with the symplectic form, then the stationary phase approximation 
is exact.  \end{prop}

\noindent{\bf Proof:} In this case we may consider the exponential 
map  at $x_0$,
$\exp: T \to N,$ which then satisfies $ f \circ \exp \, = \,
\exp \circ \, df $. (As $f$ is an isometry, it preserves geodesics.)
Of course $df \, \in \, U(T)$, the unitary group of the tangent 
space. 
The 2-form $\exp^\ast  \omega $ is a symplectic form on $T$ (the tangent
space viewed as a manifold rather than as a vector space): this 
symplectic form is preserved by $df$. 
Darboux's theorem 
thus says there is a diffeomorphism $\alpha: \, T \to T$
such that $\alpha^\ast  \,\exp^\ast  \omega  \, = \, \omega_0$, a 
 differential form 
with constant coefficients in the coordinates on  $T$.  Moreover,
$\alpha$ commutes with $df$ (by the equivariant version of Darboux), so 
also $f \circ (\exp \circ \alpha) = (\exp \circ \alpha) \circ \, df $. 

Thus we may replace paths $\ga(t)$ in $N$ satisfying 
$$f \ga(t) = \ga(t+1)$$
by paths $\psi $ in $T$ satisfying $$df \, (\psi(t) ) = \psi(t+1)$$
under the identification $\ga = (\exp \circ \alpha) \psi$. A strip  from 
the constant path 
$\ga_{x_0}$ to $\ga $ is provided by $u (t,\tau) \,= \, 
(\exp \circ \alpha) (\tau  \psi(t)).$ The action thus becomes 
$$ S(\ga) = S(\ga_{x_0}) -  \int \,
\omega_0(\tau \dps(t), \psi(t)) \, dt \, d\tau, $$
 which is precisely quadratic: thus the stationary phase approximation
to the path integral is exact. $\square$

Using the time dependent   metrics given by the 
complex structures satisfying (\ref{jtimed}), 
one may define a metric on $T_\ga \Om_f$ by 
\beq \label{metomf}
\langle \xi, \eta \rangle = \int_0^1 \om\left (\xi(t), J_t \eta(t) 
\right ) \, dt. 
\eeq 
We may use the metric to transform the quadratic form $\nb^2 S$ 
into a differential operator:
$$ \half \nb^2 S(\xi, \eta) = \langle \xi, D_0 \eta \rangle, 
$$
where 
\beq D_0 = - \half J_t \frac{d}{dt}. \eeq

\noindent{\bf Notation:} 
\begin{itemize} 
\item $T_{x_0} N$ will be denoted by $T$. 
\item $\onebl $ \hfill  
$\Om^0  \onebl \eqdef \onebl \{ \ga: S^1 \to N \}, $ \hfill $\onebl$
\item $\onebl$  \hfill $T_\ga\Om^0  \onebl \eqdef \onebl \{ 
\; \xi \in \Ga (\ga^\ast TN) \; \} . $ \hfill $\onebl$
\item In particular, if $\ga$ is the constant loop at $x_0$,  we have 
$$T_{x_0} \Om^0 \onebl \eqdef \onebl \{ \xi: S^1 \to T_{x_0} N \}. $$
\end{itemize}
Suppose $(df)_{x_0} $ is in the image of 
the exponential map on $\liesp(T_{x_0} N)$: i.e., we
assume there is $E \in \liesp (T_{x_0 } N) $ such 
that $df = \exp E$. (This is not always 
true, but we can always find such an $E$ in 
$\liesp (T_{x_0}N) \otimes \CC$: see \S \ref{constgen}.) 
Then we may write a path 
$$\rho (t) = \exp t E\in Sp(T),$$ such that 
$\rho(0) = 1 $ and $\rho(1) = df$. 

Using this path, we may transform our operator 
to an operator on the space $T_{x_0} \Om^0$ of {\it 
closed } loops in $T_{x_0} N$, as follows:

\begin{lemma} \label{constloops}
The spaces $T_{x_0}\Om^0$,  $T_{x_0} \Om_f$ are identified by the 
map $B: T_{x_0} \Om^0 \to T_{x_0} \Om_f$ which takes 
$\eta \in T_{x_0} \Om^0 \to \xi \in \txomf$, where
$$\xi(t) = \rho(t) \, \eta(t). $$
The natural path of time dependent complex structures on 
$T_{x_0} N$ is 
$$J_t = \rho(t) \,J \,{\rho(t)}^{-1} ,$$ 
for some fixed complex structure $J$. 
So our operator $D_0 =  - (1/2) J_t \ddt  $ transforms into 
$D:  \txom \to \txom$, where 
$$D = B^{-1} \, D_0 \,  B, $$
 so 
\beq \label{eq:ddef}
 D =  - \half  J ( \ddt + E) .   \bla \square \eeq
\end{lemma}

We shall want to expand $S$ beyond the quadratic order, 
and for  this purpose we need to identify  $\Om_f$ with 
a linear space. 
One approach can be used
when $f$ is the integral of a Hamiltonian flow, which 
generalizes the  technique used in Lemma \ref{constloops}. 
Let $H_t$ be a Hamiltonian function, which may 
in general depend on $t$. 
Observe first that if $f_t$ is a path of symplectic 
diffeomorphisms, then the Hamiltonian vector 
field  $X_\htil$ corresponding to 
the Hamiltonian function $\htil_t  = H_t \circ f_t $ is: 
$$X_{\htil_t} = (f_t^{-1})_\ast \: X_{H_t}. $$

Suppose we have a  path of symplectic diffeomorphisms 
$f_t$ corresponding to the Hamiltonian $H_t$:
\beq \label{eq:ham}
\frac{d}{dt}\, f_t(x) \,= \, (X_t)_{f_t(x)}, \eeq 
where $X_t $ is short for $X_{H_t}$. 
In this case we replace paths $\ga $  $\in \Om_f$ 
by closed paths $\psi$ $\in \Om^0$, under the identification
\beq \label{chvarb} \ga(t) = f_t \exp \, \psi(t). \eeq
(Here, \lq\lq exp" denotes the exponential map on the tangent 
space $T_{x_0} N$, with respect to the (time independent)
metric induced by the fixed complex structure $J$ introduced 
in Lemma \ref{constloops}.)
A homotopy from  the constant path $\ga_{x_0}$ to $\ga $ is
given by 
$$u(t,\tau) = f_t \exp  (\tau \psi(t) ). $$
The action is thus 
\begin{eqnarray*}
 S &=& - \int_{{\rm Im } \, u} \omega \\
 &=& - \int_{t,\tau = 0}^1 \omega_{u(t,\tau)} \,(u_\ast \ddt \, ,
u_\ast \ddu) \, d\tau \, dt                
\end{eqnarray*}
Now 
\begin{eqnarray*}
u_\ast \, \ddt &=&  \, (f_t)_\ast \exp _\ast \tau \dps + 
\dot{f_t}(\exp \tau  \psi(t))  \\
&=& \, (f_t)_\ast \exp_\ast  \tau \dps  +
(X_t)_{(f_t \exp \tau  \psi)} \\
&=& \, (f_t)_\ast \left \{  \exp _\ast  \tau \dps  +
(\tlxt)_{ \exp \tau  \psi}  \, \right \}. 
\end{eqnarray*}
where 
${\tlxt } \eqdef  { X_{H_t \circ f_t} }.   $ 
Thus we get 
\begin{eqnarray*}
S &=& - \int_{t,\tau =0}^1 \,   \omega  (\exp_\ast \tau \dps  
, \exp_\ast \psi)  \, d\tau  \, dt  \\
\bla &\onebl& \bla -  \int_{t,\tau =0}^1 \,   \omega  (
(\tlxt)_{\exp \tau \psi}, \exp_\ast \psi)  \, d\tau \, dt.  
\end{eqnarray*}
The first term we recognize as 
minus the logarithm of the holonomy of \call \ around the
closed path $\exp \psi(t)$. The second term is 
\begin{eqnarray*}
- \int_{t,\tau = 0}^1 &\,& (\iota_{\tlxt } \omega)_{\exp \tau \psi} 
\left \lbrace
\frac{d}{d\tau} (\exp \tau \psi) \right \rbrace  \, d\tau \, dt  \\
\onebl &=& - \int \, (d\htil_t)_{\exp \tau \psi} 
\left \lbrace  \frac{d}{d\tau} 
\exp (\tau \psi)
\right \rbrace \, d\tau  \, dt  \\
&=& - \int_0^1 \, \htil_t(\exp \psi(t) ) \, dt. 
\end{eqnarray*} 
(We assume $H_t (x_0) = (dH_t)_{x_0}=0 $ for all $t$; the same
is thus true for $\htil_t$.)
Thus  the action is:
\beq g
\label{eq:hamact}
\exp (iS) = {\rm Hol} \,(\exp \psi ) \, \exp \left \lbrace
-i \int_0^1 \htil_t(\exp \psi(t) ) \, dt \right \rbrace . \eeq

\noindent{\bf Remark:} If $H $ is {\it independent} of $t$, then it is 
conserved under the Hamiltonian flow, so $\htil = H$. 

\noindent{\bf Remark on Jacobians:}
 Formally, the path integral measure 
on $\Om_f$ is the \lq\lq Liouville 
volume" 
$$\Bigl (\cald \gamma\Bigr )_\ga
 = \prod_t \, d \mu_{\ga(t)} , $$
where $d \mu$ is the symplectic volume on $N$. This is 
invariant under the transformation 
$$[\psi(t)] \in \Om^0 \to [f_t \psi(t)]
\in \Om_f,$$ 
so the change of variables (\ref{chvarb})
does not formally produce a Jacobian in the 
path integral measure. 

\subsection{Cartan subgroups}   \label{s:cartan}
The leading order term of the stationary phase approximation for the
partition function will depend on the value of 
$df$ at fixed points $x_0$. Actually it will only be  finite for 
those values of $df$ in $Sp(2n, \RR)$
 which do not have $1$ as an 
eigenvalue.  
We analyze this leading order term by decomposing $Sp(2n, \RR)$ 
into its Cartan subgroups. 
(See \cite{b:Aed},
\cite{b:ASch} for general material on 
Cartan subgroups and subalgebras.) A {\it Cartan subalgebra} 
(CSA) $\frak{b}$
of the Lie algebra $\frak{g}$  of the Lie group $G$ is a maximal abelian 
subalgebra consisting of semisimple elements. %
These are obtained as the centralizers of almost every $X \in \frak{g}$.
The corresponding {\it Cartan subgroup}  is $B = Z_G(\frak{b}) $, the
centralizer in $G$. 
For compact groups, there is only one Cartan subalgebra up to 
conjugacy, the Lie algebra of the maximal torus. For noncompact groups there
are a finite number of conjugacy classes of Cartan subalgebras. 

The stationary phase approximation to our 
partition function $Z$ will be finite if 
$1$ is not an eigenvalue of $df$ for any 
fixed point $x_0$. If 
$U$ is  a {\it regular}   element of $G = Sp(n, \RR) $ 
(i.e., if its centralizer $Z_G(U)$  in $\frak{sp}(n, \RR) $  consists 
of semisimple elements and 
has minimal dimension), then
$U$ is contained in a unique Cartan subgroup, namely  $Z_G(U)$. 
We 
restrict ourselves to considering those $df$ that are in certain Cartan 
subgroups that are representatives of the conjugacy classes of Cartan 
subgroups. 
\begin{prop} \label{blkdec} \cite{b:Sug}
Representatives of the conjugacy classes of Cartan subalgebras
of $\frak{sp} (2n,\RR) $
are given by block diagonal matrices in blocks of the following forms:
$$ \begin{array}{l} \mbox{\bf (a)}: \phantom{aaaa} \end{array}
\hfill 
  \left \lbrack \begin{array}{cc} 0 & -h \\ 
                                   h & 0 \end{array} \right \rbrack
\begin{array}{c}  \phantom{aaa} (h \in \RR)  \end{array} 
\hfill \bla    $$
$$  
\begin{array}{l} \mbox{\bf (b)}: \phantom{aaaa} \end{array}
  \left \lbrack \begin{array}{cc} h & 0 \\ 
                                   0 & -h \end{array} \right \rbrack
\begin{array}{c}  \phantom{aaa} (h \in \RR)  \end{array}    $$
\beq \label{eq:cartan}  \phantom{aaa} \eeq
$$
\begin{array}{l} \mbox{ \bf(c)} : \phantom{aaaa} \end{array}
  \left \lbrack \begin{array}{cc} A & 0 \\ 
                                   0 & -A^T \end{array} \right \rbrack,
\phantom{aaa} A = 
\left \lbrack \begin{array}{cc} h_1 & -h_2 \\ 
                                h_2    & h_1 \end{array} \right \rbrack
 = h_1 + i h_2 = z \in \CC.  \bla  $$
\end{prop}
This decomposition 
corresponds to the fact (\cite{GS}, p. 116)
that if $z \in \CC$ is an eigenvalue 
of $X \in \frak{sp}(2n, \RR)$ then so are $-z$ and $\bar{z}$: so 
eigenvalues come in pairs $\pm  h,  \pm i h$ \ ($h \in \RR$) or
4-tuples $\pm z, \pm \bar{z}.$ The sets of block diagonal 
matrices with different 
possible orderings of the
blocks are conjugate.

\noindent{\bf Examples.} 
\begin{itemize}
\item
$G = Sp(2 , \RR) = SL(2, \RR).$ There are two
Cartan subalgebras; all regular elements of $\frak{g}$ are conjugate to 
an element of type  $(a)$ (i.e. $\frak{u}(1)$: {\it unitary}) or to 
one of type $(b)$ ({\it hyperbolic}). 
The Cartan subgroup corresponding to 
the hyperbolic case is 
$$  \left \lbrack \begin{array}{cc} \alpha & 0 \\ 
                               0 & \alpha^{-1} \end{array}
\right \rbrack , \alpha \in \RR^\times $$
which has two components.   

The  elements of $SL(2, \RR)$ 
which are not in a Cartan subgroup are the {\it parabolic}
elements, namely those conjugate to 
$$\pm \matr{1}{\al}{0}{1}. $$
The maximal abelian subalgebras corresponding to these 
elements consist of {\it nilpotent} 
elements rather than semisimple ones. 

The Cartan subgroup containing $U \in SL(2, \ZZ) $ is determined 
by the trace $\Tr(U)$: i.e., 
$$U \onebl {\rm  is} \begin{cases} {\rm hyperbolic} , & |\Tr(U)| >2; \\
{\rm unitary}, & |\Tr(U)| <2; \\
{\rm parabolic},  & |\Tr (U )| = 2. \end{cases}. $$

\item 
$G = Sp(4, \RR)$. There are four conjugacy classes of CSA:
in the notation of (\ref{eq:cartan}), these are 
$$\left \lbrack \begin{array}{cc}{\bf(a)} & 0 \\
                                  0 & {\bf(a)}  
    \end{array} \right \rbrack,
\left \lbrack \begin{array}{cc}{\bf (a)} & 0 \\
                                  0 & {\bf (b)} 
     \end{array} \right \rbrack ,
\left \lbrack \begin{array}{cc}{\bf (b)} & 0 \\
                                  0 &{\bf  (b)}   
   \end{array} \right \rbrack $$
and ${\bf (c)}$. 

\end{itemize}

\subsubsection{Reduction to constant loops: the general case} \label{constgen}
The disconnectedness of the
hyperbolic Cartan subgroup of $SL(2, \RR)$ is all
that prevents elements of 
$Sp(2n, \RR)$ from being in the image of the 
exponential map on the Lie algebra. As it stands, our  construction
of the operator $D$  (Lemma \ref{constloops})  is only valid
for $df = \exp \hatx$, $\hatx \in \frak{sp}(2n,\RR)$. 
To extend it to the 
general case, it suffices to find a 
way to treat 
$$df = - \matr{e^h}{0}{0}{e^{-h}}, \: h \in \RR. $$
This is done as follows. The operator $D$ defined in Lemma 
\ref{constloops} acts
in the first instance on $L^2 (S^1, \RR^{2n})$; however, we can compute
its determinant and eta invariant from its action on $L^2 (S^1, \CC^{2n})$. 
In the case when $df$ is not in the identity component of its Cartan subgroup,
it still makes sense to   define $D$ as an operator acting on 
$L^2 (S^1, \CC^{2n})$ using the formula (\ref{eq:ddef}) but where $\hatx$
is a {\it complex} matrix such that $df = \exp \hatx$: we take, for 
the above $df$, 

$$ \mbox{(\ref{eq:cartan}){\bf (d)}\hfill  } \bla 
\hatx = \matr{h+ i \pi}{0}{0} {-h+ i \pi}. \hfill \bla  $$
Notice that for $$J = \matr{0}{-1}{1}{0}, $$
the endomorphism $J \hatx$ is self-adjoint 
on $\CC^{2n}$; this is the reason for 
making this particular choice of $\hatx$.

\subsection{Calculation of det {\it D}: absolute value} \label{s:det}
We need the determinant of the quadratic form from (\ref{eq:ddef}) 
$$D = -\frac{J}{2} (\tder + \hatx) $$
on $L^2(S^1, T \otimes \CC)$, where $ df = \exp \hatx$. 
We observe:

\begin{lemma}
$J \hatx = - \hatx^\ast J $, where $\ast$ is 
the adjoint with respect to the metric 
$g$ on $T = T_{x_0} W$ compatible
with $J$ and $\omega$.  The operator $D$ is self-adjoint.
\end{lemma}
\noindent{\bf Proof:} 
$J^\ast = - J$, 
and 
 $\tder^\ast = - \tder$ by integration by parts. It thus remains to 
prove $J \hatx$ is self-adjoint. For $\hatx \in {\frak{sp} }(T)$,
we have 
$$g (J\hatx  \phi, \psi) = \om (J \hatx  \phi, J \psi) $$
$$ = \om( \hatx  \phi, \psi)  = - \om ( \phi, \hatx \psi) $$
$$ = g ( \phi, J  \hatx  \psi). $$
This extends to case (\ref{eq:cartan} d)  as well, 
since there $E = E_0 + i \pi 1$ 
with $E_0 \in {\frak{sp}} (T). $  \hfill $\square$ 

We shall formally compute 
$|\det D|$ by multiplying together all the eigenvalues
of $D$, as 
in \cite{b:A!circ}.

\begin{lemma} \label{detindep}
The absolute value $|\det D|$ is independent
of $J$. 
\end{lemma}
\noindent{\bf Proof:} The eigenfunctions of $D$ are the solutions $\psi$ of
$$\dps = (- \hatx + 2 J \lambda) \psi$$
for eigenvalues $\lambda$. Writing $- \hatx + 2 J \lambda = A$, these
are the functions $\exp A t$: but because they must be periodic, they must
correspond to eigenvectors of $A$ with eigenvalues $2\pi i m$
($ m \in \ZZ$), i.e.,
$$\det (- \hatx +2 \lambda J + 2 \pi i m) = 0 $$
or 
\beq \label{eq:eig}
\det J (J \hatx + 2 \lambda -  2 \pi i mJ) = 0. \eeq
The product of the $2n$ solutions $\lambda $ to this (for 
given $m$) is  
$$1/2^{2n} \det(J\hatx - 2 \pi i m J)  = 
  1/2^{2n} \det(\hatx - 2 \pi i m ),  $$
independent of $J$. \hfill $\square$

The determinant and eta invariant are multiplicative on block diagonal 
operators, so it suffices to compute the contributions to them from the 
Cartan subalgebra blocks described in \S \ref{s:cartan}. 
The contributions to $\prod_m   \vert \det (\hatx - 2 \pi i m) \vert$ 
are the following: 

\noindent{\bf (a):}  
$\hatx$ has eigenvalues $\pm ih$, so  $4 \pi^2 m^2 - h^2.$ 
 
\noindent{\bf (b):} Eigenvalues $\pm h$, so $4 \pi^2 m^2 + h^2.$ 

\noindent{\bf (c):} Eigenvalues $\pm z$, $\pm \bar{z}$, so
$(4 \pi^2 m^2 + z^2 )(4 \pi^2 m^2 + {\bar{z}}^2).$ 

\noindent{\bf (d):} Eigenvalues $\pm h + i \pi$, so 
$h^2 + \pi^2 (2m-1)^2$. 

\noindent  This is the contribution for the block from each $m \in \ZZ. $

To make sense of the infinite product over all $m$, we recall
$$ \sin \, z = z \prod_{n=1}^\infty  \left  ( 1 - \frac
{z^2}{\pi^2 n^2} \right   ), $$
i.e., 
$$ \alpha \prod_{m  \ge 1} (\alpha^2 + 4 \pi^2 m^2) = 
(e^{\alpha/2} - e^{-\alpha/2} )  ( \prod_{m\ge 1} 4 \pi^2 m^2). $$
We  \lq \lq renormalize\rq \rq by discarding the factor 
$( \prod_{m\ge 1} 4 \pi^2 m^2)$.  
\begin{prop} \label{formult}
The following are the contributions to 
$|\det D |$ 
from the blocks  in (\ref{eq:cartan}): 

\noindent{\bf(a):} 
$$|\det D| = h^2 \prod_{m \ge 1} (4 \pi^2 m^2 - h^2)^2 = 4 
|\sin (h/2)|^2$$

\noindent{\bf(b):} 
$$|\det D| = h^2 \prod_{m \ge 1} (4 \pi^2 m^2  +h^2)^2 = 
|e^{h/2} - e^{-h/2} |^2         $$

\noindent{\bf(c):} 
\begin{eqnarray*}
|\det D| &=& z^2 {\bar{z}}^2 \prod_{m \ge 1} 
(4 \pi^2 m^2  +z^2)^2 \:(4 \pi^2 m^2  + {\bar{z}}^2)^2 \\   
&=& |e^{z/2} - e^{-z/2} |^2 |e^{\zbar/2} - e^{-\zbar/2} |^2.  
\end{eqnarray*}

\noindent{\bf (d):} 
$\vert \det D \vert  = (e^{h/2} + e^{-h/2})^2. $ \hfill  $\square $
\end{prop}

In the above (a - c), 
a factor $( \prod_{m\ge 1} 4 \pi^2 m^2)^2$ has been discarded from each
block of dimension $2$, two from a block of dimension 4.  
Case (d) follows 
using the analogous formula 
$$\cos z = \prod_{n=1}^\infty  \left ( 1 - 4  \:
\frac{z^2}{(2n-1)^2 \pi^2  } \right )
$$
and discarding a factor 
$\prod_{n \ge 1}\{ \pi^2 (2n-1)^2    \}^2. $ 

\begin{rem}{\bf : Zeta function determinant calculation}

These values may also be obtained by observing that 
$$ | \det J(\tder + E ) | = |\det (\tder + E) |, $$ 
and then diagonalizing $E$ and using the zeta function 
determinant $|\det (\tder + \la)| $ for $\la \in \CC^\ast $.  
Zeta function determinants of this type are 
treated in \cite{forman}.
\end{rem}

We see:
\begin{prop}
$|\det D |^{-1/2} $ is the absolute value of the 
character of the metaplectic
representation of $df$.          $\square$                            
  \end{prop}

\subsection{Calculation of the eta invariant} \label{s:eta}

For our $D$, $\eta(D)$ will in general depend on the complex 
structure $J$. A motivation for this is provided by the calculation 
in  Lemma \ref{detindep}. 
$J$ is traceless since it is
in $\frak{sp}(2n, \RR)$; 
thus using (\ref{eq:eig}), the eigenvalues corresponding
to a given $m$ sum 
to $-\frac{1}{2} {\rm Tr} (J\hatx)$, which depends on $J$. 
The eta invariant measures this asymmetry.  

\subsubsection{The {\it SL(2, R)} case}  \label{sltwor}
We may confirm  this in the case of 
$SL(2, \RR)$ by the following calculation. 

The 
computation is easy for $SL(2, \RR)$. We start with the calculation 
(e.g. \cite{b:Gilkey}, p. 82) for the eta invariant of the operator 
$A=i\tder + a \; (0 <a <1)$  on $L^2(S^1, \CC)$, whose spectrum is
$n + a \; (n \in \ZZ)$; it is 
\beq  \eta(A) = 1-2a,  \bla \mbox{$0 <a < 1 $}. \eeq 
We seek the eta invariant of the operator $D$ from (\ref{eq:ddef}).
We write $D = -(J/2)(\tder + \hatx)$ 
Since $\eta (D)$ is invariant under simultaneous
conjugation of $J$ and $\hatx$, we may 
 fix $J$ and compute 
 $\eta(D)$ as a function of $\hatx$.  We take 
$$ J = \left \lbrack \begin{array}{cc} 0 & -1  \\
                                       1 & 0  \end{array} \right \rbrack$$
 and 
$$\hatx = \sigone \matr{0}{1}{1}{0} + \sigtwo \matr{0}{-1}{1}{0}
+ \sigthr\matr{1}{0}{0}{-1}, $$
in terms of real coefficients ${\Sigma}_i$.
The eigenvalues of $D$ are then
\beq \label{eivalsl}
\lambda_m^{\pm} = \frac{1}{2} \left (  \sigtwo \pm (\sigone^2
 + \sigthr^2 + 4\pi^2 m^2)^{1/2}  \right ) \eeq
(two eigenvalues for each $m \in \ZZ $). The overall scaling of the 
eigenvalues does not affect the eta invariant, so replacing 
$\tilde{\sigma_i}$ by $\sigma_i = \tilde{\sigma_i}/(2\pi) $, 
we may use
\beq \label{lapm}
\lambda_m^{\pm} =  
\sigma_2 \pm \left ( \kappa^2 + m^2 \right )^{1/2},
\eeq
where we have defined $\kappa^2 = {\sigma_1}^2 + {\sigma_3}^2$. 

For $\kappa = 0 $, the eta invariant is just the same as that of $A$ above
(multiplied by 2 since we have twice as many eigenvalues), i.e., 
$\eta = 2(1- \sigtwo/\pi )$. 
We differentiate with  respect to $\kappa$  
 to get 
$$\frac{\partial}{\partial \kappa}\, \eta (s, \kappa) \,= \,   
\mp s \sum_\pm \sum_{m \in \ZZ} \:\frac{1}{| \sigma_2 \pm 
\sqrt{\kappa^2 + m^2} |^{s+1} }  \: 
\frac{\kappa}{\sqrt{\kappa^2 + m^2} } . $$
This is absolutely convergent: for large $m$ and $s$ near zero, 
it goes like
$ s \sum_m (\kappa^2 + m^2)^{-1} $.  
Now $\eta(s, \kappa)$ 
(for all $s \ge 0$)
is $C^\infty$ in $\kappa$, except at 
those values of $\kappa$ where some $\la_m^\pm (\kappa)$ 
becomes zero. Even at  these  values, $\eta$ is 
$C^\infty$  as a function into $\RR/\ZZ$: in fact, the quantity 
\beq \tilde{\eta}(D) = \frac{\eta(D) + h (D) }{2} \eeq
is $C^\infty$ as a function into $\RR/\ZZ$.
Here, we have introduced the notation 
\beq h(D) = \dim \Ker (D). \eeq  
Thus, away from the zero eigenvalues, we may interchange 
$\frac{\pr}{\pr \kappa}$ with taking the limit as $s \to 0$
to conclude
$$\frac{\pr}{\pr \kappa} \tilde{\eta} (0, \kappa) = 0, $$
i.e., provided $0 < \sigtwo <2 \pi$, we have  
$$2\tilde{\eta}(0, \kappa) 
= 2 - 2 \tilde{\si_2}/\pi \bla ({\rm mod} 2\ZZ). $$
For small enough $\kappa $, this is the exact value  (not just 
mod $\ZZ$). 
Notice also that  ${\rm Tr }J\hatx = -2 \sigtwo$, confirming
the remarks at the beginning of this section.  

\subsubsection{The general case}
In general 
$$D = - \half J (\tder + E) = \; \; - \half 
\left ( J \tder 
+ \half  (JE + EJ) + \half (JE-EJ) \; \; \right ). $$
Now $JE + EJ$ commutes with $J$ while $JE-EJ$ anticommutes
with it. So we may apply a theorem 
due to  Cheeger \cite{Cheeger}
(Theorem \ref{cheeger} below), which implies  that adding  
$JE - EJ$ does not change $\tilde{\eta} \pmod{\ZZ}$. 
We thus need to evaluate $\tilde{\eta}(D_0)$, where 
$$D_0 = - \half \left ( J \tder + \half  (JE + EJ) \right ). $$
Now we may diagonalize the restriction of $JE+EJ$ to the 
$+i $ eigenspace of $J$: suppose it has eigenvalues $\la$. 
Then 
\beq \frac{\eta + h}{2} (D_0)  
= - 2 \sum_\la \frac{\eta + h}{2}  \left ( i \tder + \la \right ) 
\onebl \pmod{\ZZ}  , \eeq
$$= - \sum_\la \left (1 - 2 \frac{\la}{2 \pi} \right ) = 
\frac{2}{2 \pi } \Tr_{T_{x_0}'N}
 (JE)  \onebl \pmod{\ZZ} , $$
$$ = \frac{\Tr JE} {2 \pi} \pmod{\ZZ}. $$
Since we have assumed $\ker D = 0, $ this gives 
\begin{lemma}
The eta invariant of $D$ is given by 
$$\eta(D) = \frac{\Tr (JE)}{\pi} \;\; \pmod{2 \ZZ}. \bla \square $$
\end{lemma}

\subsubsection{Choices of complex structures}
To  make sense of the dependence of the eta invariant on $J$, we 
note that there is a canonical 
sensible choice of complex structure for each of the blocks in the Cartan 
subalgebra (see \S \ref{s:cartan}), i.e.,

\beq \label{stdcplx}
 {\bf (a), (b), (d)}:\; J = \matr{0}{-1}{1}{0}, \phantom{aaa} 
{\bf(c)}:\; 
J= \matr{0}{-1_2}{ 1_2}{0}.  \eeq
With these complex structures, the eigenvalues of  $D$ are:
$${\bf (a)}: \hfill  \; h/2 \pm  \pi m\hfill \bla $$
$${\bf (b)}: \hfill  \; \pm |h/2- i \pi m |\hfill \bla $$
$${\bf (c)}: \hfill \; \pm |z/2 -  \pi i m|, \pm |\bar{z}/2 -  \pi i m |
\hfill \bla  $$
$${\bf (d)}: \hfill  \; \pm | h/2 - i \pi (m - 1/2) |. \hfill \bla 
     $$
With these complex structures, blocks of type $(b)$ , $(c)$   and $(d)$
contribute nothing to the eta invariant, 
because the eigenvalues are  in pairs $\pm \la$. 
 For case $(a)$, we get
\beq \label{acase}\eta = 2 - 2h/\pi \bla (0 < h < 2 \pi ).\eeq
Thus to define the eta invariant for arbitrary $\hatx$, we use this canonical
complex structure on a block diagonal Cartan subalgebra containing a 
conjugate of $\hatx$. 

Given $df$, one may associate a {\it class} of complex
structures to it by choosing $g \in Sp(2n, \RR)$ such
that $df = g^{-1}B_0 g$, where $B_0$ is one of our standard 
forms for elements of Cartan subgroups; one then defines
the associated standard complex structure
$J_0$  from (\ref{stdcplx}), and $J = g^{-1} J_0 g$. 
For instance, if $df$ preserves a complex 
structure, this prescription says to use   the 
complex structure it preserves 
to define the eta invariant.   In this
unitary case, the complex structure is in 
fact unique, provided $df $ has  no eigenvalues
$\pm 1$. However,  if there are hyperbolic blocks 
one may not in general associate a unique
complex structure by this means. For instance,
conjugating by the hyperbolic Cartan 
subgroup of $SL(2, \RR)$ fixes
any $df$ in this subgroup, but 
transforms  our standard complex 
structure  as follows:
$$J_0 = \matr{0}{-1}{1}{0} \onebl \mapsto 
\onebl J = \matr{0}{- \mu^{-1} }{\mu}{0} \onebl, \mu >0 . $$

 The eta invariant is  well defined by these choices of complex 
structures:
it is determined by the eigenvalues of $\hatx$ of the form $ih, \; h \in \RR$.
The only ambiguity is thus the sign of $h$. However, if we could 
change the sign of $h$ by conjugation, we could conjugate 
the complex structure $J$ to $-J$, where
$$ J= \matr{0}{1}{-1}{0}.$$
 This is impossible since conjugation of a positive complex structure
(one for which the associated symmetric quadratic form is positive 
definite) produces another positive complex structure. Thus there is a 
canonical way to define the eta invariant of $D$, despite the complex 
structure dependence.

A more sophisticated  procedure to deal with the complex structure
dependence of the eta invariants is discussed in 
\S \ref{regul}.

\subsubsection{The holomorphic Lefschetz formula} \label{hollef}
Combining the value (\ref{acase}) 
of $\eta$ (for $h \in (0,2 \pi)$) with the value
$4 \sin^2 (h/2)$ for $\det(D)$, we  have 
$$ |\det D|^{-1/2} \, e^{ \frac{i \pi}{4} \eta(D) }  = \frac{1}{1 - e^{ih} }.
$$   
If  $df_{x}$ is conjugate to a unitary matrix at all 
fixed points $x$, this gives 
\beq \label{holpoint}
Z = \sum_x \: \frac{{\rm Tr} \tilde{f_x}}{\det_\CC(1 - df_x) } \eeq
for the leading order contribution to the partition 
function from ${x_0}$. 

If $f$ preserves
not only the symplectic structure but also a compatible holomorphic
structure on $W$, and if $f$ lifts to a holomorphic map $\tilde{f}$
on $\call$,  then $\tf$ acts also on the sheaf cohomology
groups. 
The {\it holomorphic Lefschetz fixed point 
formula } (see \cite{b:AB!LF})  then states that 
\beq \label{eq:lefsch}
\sum_i (-1)^i  {\rm Tr} \, {\tilde{f}}_{H^i (N, \call)}
 = \sum_P \frac{ {\rm Tr} \,{\tilde{f}}_{(\call_P)}}
{\det_\CC (1 - df_P) } , 
\eeq
where the right-hand sum is over the fixed points $P$ of $f$. 
The right hand side is precisely the  sum of the contributions 
from (\ref{holpoint}) over all fixed points.  Furthermore, 
in this situation, we have shown above (Proposition\ \ref{stexact}) 
that the partition function is given precisely by the leading
order term.  Thus if $f$ preserves a holomorphic structure, we have 
\beq  \label{holr}
Z(M,k) = 
\sum_i (-1)^i  {\rm Tr} \, {\tilde{f}}_{H^i (N, \call^k)}. \eeq

If $H^i$ vanishes for $i > 0$, this suggests an 
interpretation in terms of the quantization of 
the theory. In geometric quantization, one 
way to quantize a symplectic manifold $(N,\om)$ is 
to pick a  prequantum line bundle $\call$, which acquires a 
holomorphic structure given a compatible complex
structure $J$ on $N$. The physical 
Hilbert space $\calh$ is then defined as 
$$\calh = H^0(N, \call). $$
From this point of view, the holomorphic Lefschetz
formula appears as the statement that our path integral 
evaluates the trace of $f$ on the physical 
Hilbert space. 
If the higher cohomology groups do not vanish, it appears 
from our result (\ref{holr})
that the physical Hilbert space should be interpreted
as a virtual vector space $\calh = \oplus_i (-1)^i H^i(N, \call). $

\section{Relation between SQM and Chern-Simons} \label{relscs}

The following properties are obvious, and are in any 
case discussed in \cite{DS} (\S  3):

\begin{prop}
The space of connections on $P_\be$ is given by 
$$\cala(P_\be  ) = \left \{ \atil = A(t) + \Phi (t)\, dt \mid 
\tlbe^\ast A(t+1)  = A(t), \tlbe^\ast \Phi(t+1) = \Phi(t) \right 
\}, $$
where 
$$\Phi : \:\RR \to \Om^0(\Si, \ad P), \bla A: \: \RR \to \Om^1(\Si, 
\ad P) $$
are $C^\infty$ . 
The space of gauge transformations on $P_\be$ is 
$$\calg (P_\be) = \left \{ g: \RR \to \calg(P) \mid 
\tlbe^\ast g(t+1) = g(t)\right \} , $$
where $g$ is also $C^\infty$ from $\RR$ to $\calg(P)$.  $\bla \square$
\end{prop}

The action of a gauge transformation $g$ on $\atil$ is 
$$g^\ast \atil (t) = g(t)^\ast A(t) + (g^{-1} \dot{g} + 
g^{-1} \Phi g) dt. $$
In this language the Chern-Simons functional becomes 
(\cite{DS}, (3-2)) 
\beq \label{3.2:cstime}
\cs(A + \Phi dt) = \frac{1}{4 \pi^2} \int \Tr (F_A \wedge \Phi) dt 
- \frac{1}{8 \pi^2} \int \Tr\Bigl ( (A - A_0) \wedge \dot{A} \Bigr
) \, dt. 
\eeq
One thus sees in particular that 
\begin{lemma} \label{3.2:csgauge}
\begin{enumerate}
\item If $\atil$ satisfies $F_{A(t)} = 0$, then so does 
$g^\ast \atil$. In other words, gauge transformations
preserve paths in the space of flat connections. 
\item For paths $A(t)$ with $F_{A(t)} = 0$, the variation of the 
Chern-Simons functional 
under an infinitesimal gauge tranformation
$A(t) \to A(t) + d_{A(t)} a(t)$ is zero.  \qqed 
\end{enumerate}
\end{lemma}
The second part follows because one can rewrite  the 
variation of the Chern-Simons functional as (see 
\cite{DS}, after (7-4)):
$$\delta CS_{A + \Phi dt} (\al + \phi dt) = \int_\Si \Tr
\Bigl \{ \, (\dot{A} - d_A \Phi) \al + F_A \phi \,\Bigr \} dt. $$
One then integrates 
by parts and uses the fact that $d_A \dot{A} = \pr F_{A(t)} /\pr t = 0. $

Now the symplectic form on $\cala (\Si)$ is chosen to be 
\beq \label{3.2:syma}
\om(a,b) = \frac{i}{2 \pi} \int_\Si \Tr (a \wedge b). \eeq
The normalization is chosen
so that $\int_\Si\om \in 2 \pi\ZZ.$ As above,  this corresponds 
 to the connection form 
\beq \label{3.2:conncal}
\Theta_A(a) =  \frac{1}{4 \pi} \int_\Si \Tr((A - A_0) \wedge a).\eeq 

What follows is the formal proof that the partition function of 
symplectic quantum mechanics is equal to the partition function 
of Chern-Simons gauge theory.

\begin{prop} \label{chsqfo}
The Chern-Simons path integral (\ref{ch2:zpath}) over 
the space $\cala (\Si_\be)$
equals the SQM path integral corresponding to 
the symplectic manifold $N = \calm (\Si)$ with the symplectic
diffeomorphism $f$ induced by $\be$. 
\end{prop}

\noindent{\bf \lq\lq Proof'' (formal):}
Let us examine the result of inserting the 
expression (\ref{3.2:cstime}) into the Chern-Simons
path integral. One first performs the path integral 
over $\Phi(t)$ for all t: this imposes the 
condition $F_{A(t)} = 0.$ Thus the partition function
reduces to an integral over paths in the space of 
{\it flat} connections on $\Si$; the Lagrangian on 
these paths is $-(1/4 \pi) \int \Tr( (A-A_0) \wedge \dot{A}) \, dt. $ 
Because of Lemma \ref{3.2:csgauge}, the path integral descends
to a path integral over paths in the moduli space $\calm$ 
satisfying the appropriate endpoint condition.

Thus, after doing the $\Phi$ integration, the Chern-Simons path 
integral reduces to the following integral over paths $A(t)$ 
in the space of flat connections on $\Si$: 
$$\int e^{ - i k \int  \theta_A (\dot{A}) dt } \, \cald A(t), $$
which is precisely the SQM partition function 
for paths in $\calm$ with the endpoint condition imposed by 
$f$. \qqed

\section{Large {\it k} limits in Chern-Simons and 
SQM} \label{llim}

This section provides evidence 
for the formal argument given in  
Proposition\  \ref{chsqfo}. For an arbitrary surface $\Si$ with a 
diffeomorphism $\be$, we show that the critical points of the SQM
Lagrangian on $\calm$ correspond to flat connections 
on $\Si_\be$, and that the data 
from the large $k$ limit of  the Chern-Simons theory 
(the Reidemeister torsion and the Chern-Simons invariant
at flat connections) correspond precisely to the 
data from the large $k$ limit of SQM. 

\begin{lemma}
The connection $\atil = (A, \Phi)$ on $\Si_\be$ is  flat if 
and only if $A(0)$ is a fixed point of $f: \calm \to \calm.$
\end{lemma}

\noindent{\bf Proof:} 
 The curvature of $(A, \Phi)$ is $(dA + A^2) + (d_A \Phi - 
\dot{A}) dt. $ 
(More detail is given in \cite{DS}, Lemma 
3.1.) \qqed

\subsection{The Chern-Simons functional}
\begin{prop}
The Chern-Simons action $e^{2 \pi ik \cs(\atil)}$ at a flat connection
$\atil$ is equal to $\Tr \,\tlbe$, where $\tlbe$ is the lift 
of $\be $ to the prequantum line bundle. 
\end{prop}
\noindent{\bf Proof:} Recall from \S \ref{chap1} 
that if a lift of $f$ preserving the connection $\theta$ on 
$\call$ exists, it is unique up to a constant factor in 
$U(1)$. However, as we shall describe, the Chern-Simons 
functional on connections on the mapping 
torus is precisely the function that is used to 
lift the action of bundle automorphisms from 
$\cala(\Si)$ to $\call$ in a way that preserves the 
connection. For a related treatment see \cite{DS} or \cite{RSW}.

We recall (\cite{ADW}, \cite{RSW}) 
that one way to define the prequantum line 
bundle on $\calm$ is to start with the trivial bundle
$\call = \cala(\Si) \times U(1)$ over $\cala(\Si)$ 
and construct a lifting of the action of $\calg_\Si$ on
$\cala(\Si) $ to $\call$. Actually we will lift not just
$\calg_\Si$ but ${\rm Aut}(P_\Si).$ For $\psi \in {\rm Aut} (P_\Si)$, 
 the lift is  defined by 
\beq \bar{\psi}(A,z) = (\psi^\ast A, e^{2 \pi i \Theta(A, \psi)} z).
\eeq
Here,  one defines the function $\Theta$ as follows. Given 
$A \in \cala(\Si)$ and $\psi \in {\rm Aut}(P_\Si)$ covering
$\be \in {\rm Diff}(\Si),$   we pick a 
connection $\tilde{A}$ on $\Si \times \RR$ such that 
$\atil(t + 1 ) = \psi^\ast \atil(t).$  This connection 
$\atil$ thus defines
a connection on  the mapping torus $\Si_\be$. 
One defines 
$$\Theta (A, \psi) = \cs(\atil).$$
This is independent of the choice of interpolation
$\atil$, since one can get between different
interpolations by a gauge transformation on 
the closed manifold $\Si_\be$. 
One may check that the map $\bar{\psi}: \call \to \call$ 
preserves the connection. 
Restricting  to the gauge group $ \calg_\Si  \subset {\rm Aut}(P_\Si),$ 
one lifts the gauge group action to $\call$ and thus  defines a quotient 
bundle $\call \to \calm(\Si) \subset \cala(\Si)/\calg_\Si.$ Furthermore, 
if $\psi = \tlbe \in {\rm Aut}(P_\Si)$ covers $\beta \in  
{\rm Diff} \,(\Si)$, the map $\bar{\psi}$ descends to give 
a lift of $f_\be: \calm \to \calm$ to an 
automorphism $\tilde{f_\beta}: \call \to \call$  of the prequantum line
bundle over $\calm$,
preserving the natural connection.  \qqed

\begin{rem} One uses a periodic $C^\infty$ family of connections 
parametrized by  $\RR$ rather than just a family
parametrized by $[0,1]$ satisfying an 
endpoint condition, because otherwise one may fail
to obtain a $C^\infty$ connection on $\Si_\be$. 
In particular, it is  incorrect to use the linear interpolation
between $A$ and $\psi^\ast A$. $\square $  \end{rem}

\subsection{Reidemeister torsion}
\begin{lemma} The fundamental group  of the mapping torus 
$\Si_\be$ is given by 
$\pi_1 \Si_\be = \ZZ \tilde{\times} \pi_1 \Si$,
where $\ZZ$ acts on $\pi_1 \Si$ via the diffeomorphism $\be$ of $\Si$. 
In other
words, $\pi_1 \Si_\be$ is generated by $\pi_1 \Si$ 
and one new generator $\ga$, with  the additional  relation
$\ga^{-1} \si \ga = \be_\ast \si.$  \qqed
\end{lemma}

\begin{lemma} \label{repexpl} We have 
$$
{\rm Rep} (\pi_1 \Si_\be, G) = \Biggl  \{ 
(\rho, g) \mid \: \rho \in {\rm Rep}(\pi_1 \Si, G), \: 
g \in G,\:  g \rho g^{-1} = \be^\ast \rho \:\Biggr  \}.  \bla \square$$
\end{lemma}

\begin{prop}
Assume $\rho$ is an isolated fixed point of $f$ in 
$\calm(\Si)$. 
Denote by $\calm_\rho$ the subset of $\calm(\Si_\be)$ corresponding to 
$\rho$, and assume $g \in G$ is such that 
$g \rho g^{-1} = \be^\ast \rho$.  Then $\calm_\rho$ may 
 be identified with the image of the coset  $Sg$ in $G$ 
under a certain action of $S$, where the subgroup $S \subset G$ 
is the stabilizer of $\rho$.
\end{prop}

\noindent{\bf Proof:} The element $g $ must satisfy
\beq \label{3.2:gcond}
g \rho g^{-1} = \be^\ast \rho . \eeq
Thus $g \in N(S)$, the normalizer of $S$. 
All other $g'$ that also satisfy (\ref{3.2:gcond}) 
are parametrized by 
$$ g^{-1} g' \in S; $$
clearly also 
$$ {\rm Stab}(\rho,g) = S \cap Z(g).$$
Then $\calm_\rho$ is $Sg/S$, the quotient under the 
action of $S$ by conjugation (which exists because 
$g \in N(S)$).  \qqed 

\noindent{\bf Examples:}
\begin{enumerate}

\item If $\rho$ is irreducible, then $S = Z(G)$ and 
$\calm_\rho= Sg$, a set of points.

\item If $\rho$ is central, $S = G$ and $\calm_\rho = T/W$,
where $T$ is the maximal torus and $W$ the Weyl group.

\item If $G = SU(2)$, and $\rho$ is abelian but not central, 
then  $S = T.$
       \begin{enumerate}
	\item  If $g \in T$, $\calm_\rho = T$. 

\item If $g \notin T$, $S\cap Z(g)$ has dimension 0. The element 
$g$ may be chosen as $j$ (in quaternionic notation), 
and any element of $Sj$ is conjugate to $j$ via an element of $T$, 
so $\calm_\rho $ is a point. 
\end{enumerate}

\item If $\Si$ is a torus, all representations
are abelian. For a {\it generic} representation $\rho$, 
the stabilizer $S$ is the maximal torus $T$ 
and the action of $ g \in N(T)$ on 
${\rm Lie} (S)$ is just its action 
via the Weyl group $N(T)/T$. 
We may identify $Sg/S$ with the 
subspace  $(S\cap Z(g)) \,g$, on which the conjugation 
action of $S$ is trivial.  
For any fixed $g_0$, and $g \in (S \cap Z(g_0) ) \, g_0,$
we have
$$\ad (g)\,\tlbe^\ast  =  \ad(g_0) \,\tlbe^\ast \in 
\End H^1 (\Si, \rho), $$
so the induced map $df$ $ = \ad(g) \tlbe^\ast$ 
   on the tangent space is well 
defined, as  will be discussed in \S 2 of\cite{sqpaper2}.
(Note that
some of the fixed points of $f$ may be  singular points of 
$\calm(\Si).$)
\end{enumerate}

The following result is Proposition 5.6 in \cite{lenspa}, which was stated
in that paper without proof. The proof appears here.

\begin{prop} \label{proplenspa}Suppose $(\rho, g_0) $ satisfy (\ref{3.2:gcond}),
and $\rho$ is an isolated fixed point for the action of 
$f_\beta $ on $\calm(\Si)$. 
The integral over $\calm_\rho$ of the square root of 
the Reidemeister torsion is 
$$\int_{g \in S g_0/S}\,
 d \vol(g)  \:
{ \Biggl \vert  \det \Bigl (\ad(g) \tlbe^\ast - 1 \Bigr )  
\Biggr \vert^{-1/2} }. $$
Here,
 $  \ad(g) \tlbe^\ast - 1 \;\;\in \End \, (H^1 (\Si, \rho))$
is the map $df$ induced by $ \beta$ on $T_\rho \calm(\Si)$ at 
the fixed point $\rho$, and $\tlbe $ is a lift of $\be$ to 
the flat bundle $ \ad (P ) $ over $ \Si$. 
\end{prop}

\noindent{\bf Proof:} 
We seek the Reidemeister torsion of the mapping torus  $\Si_\be$ with the 
representation $(\rho,g)$ of its fundamental group, in the 
notation of Lemma \ref{repexpl}.  This is an element of 
$(\lamax H^1(\Si_\be))^2 \otimes (\lamax H^0(\Si_\be) )^{-2}$.

Using
 the obvious Mayer-Vietoris exact
sequence for the mapping torus of a surface,
we  have 
\beq \label{mayseq}
\dots \lrar H^i(\Si_\be) \lrar H^i (\Si)  \oplus H^i (\Si) \stackrel{\mu}
{\lrar} H^i (\Si) \oplus H^i (\Si) 
\lrar H^{i+1} (\Si_\be) \lrar \dots \eeq
(where the cohomology groups $H^i(\Si)$ are with respect to the differential
$d_\rho$ corresponding to the representation $\rho$, and the groups
$H^i(\Si_\be) $ are with respect to $d_{(\rho,g)}$).

We represent $\Om^i(\Si, \rho) $ by $ \left (C^i (\tilde{\Si} ) \otimes 
\lieg \otimes \CC  \right ) \, / \sim, $ where 
$\tilde{\Si} $ denotes the universal covering space of $\Si$ 
and $\sim $ denotes the equivalence relation 
given by the action of the fundamental group $\pi$ via 
the representation $\rho$. 
To define the action of $\be $ we choose a lift $\tlbe $ 
to $\tilde{\Si}$. 
We choose  the  basepoint 
in $\Si$ to be preserved by $\be$, and the lift 
$\tlbe$ is  chosen to preserve a lift of the 
basepoint.  
The map $\mu$ corresponding to the point $(\rho,g)$ is then just
$$\mu = \matr{1}{1}{1}{\ad(g) \tlbe^\ast}. $$

Using the behaviour of torsion under the 
Mayer-Vietoris sequence, 
the Reidemeister torsion we seek 
is the same as the Reidemeister torsion of 
\beq \label{torfrac}
\frac{\tau(U) \tau (V)}{\tau(U\cap V) }  = 
\frac{\tau(\Si) \: \tau (\Si) } {\tau (\Si \coprod \Si) } , \eeq
where $$\tau(\Si)\in \frac{\lamax H^1 (\Si) }
{\lamax H^0(\Si)  \; \lamax H^2 (\Si)}  
 $$
and the notation in (\ref{torfrac}) means that we take the 
determinant of the maps on the cohomology groups of $\Si$ 
arising from the Mayer-Vietoris long exact sequence.
 Clearly 
in this situation we simply want the determinant of the 
maps from the long exact sequence of cohomology groups, since 
all the scalar factors in $\tau(\Si)$ cancel between 
the numerator and the denominator of (\ref{torfrac}).

Consider the portion of the exact sequence (\ref{mayseq})
$$\dots \lrar H^1 (\Si)  \oplus H^1 (\Si) \stackrel{\mu}
{\lrar} H^1 (\Si) \oplus H^1 (\Si) \lrar \dots $$
By our assumption that $\rho$ is an isolated fixed 
point of $f_\be$, $\mu$ is an isomorphism. The contribution from 
the determinant of $\mu$ to the determinant of this 
exact sequence is then precisely 
$1/\det\left ( \ad(g) \tlbe^\ast - 1\right ). $

Consider now the portion of the exact
sequence 
\beq \label{mayseq2}
0 \lrar H^0 (\Si_\be) \lrar H^0 (\Si)  \oplus H^0 (\Si) \stackrel{\nu}
{\lrar} H^0 (\Si) \oplus H^0 (\Si) \lrar H^{1} (\Si_\be) \lrar 0 . \eeq
The space $H^0(\Si)$ may be identified with ${\rm Lie}(S)$.
The map $\nu$ is then
$$\nu = \matr{1}{1}{1}{\ad(g)}  \onebl \onebl 
\in \End \,{\rm Lie }(S),$$ 
Clearly then $H^0(\Si_\be) \cong {\rm Lie}(S) \cap Z(g)$,
the subspace fixed by the adjoint action of $g$. Since 
$\nu$ is self-adjoint in the metric on ${\rm Lie}(S)$,  
$H^1(\Si_\be)$ identifies naturally 
with $({\rm Im} \: \nu)^\perp \, \cong {\rm Ker }\,\nu.$
The standard metric on ${\rm Lie} (S)$ (from the 
basic inner product on $ \lieg$) gives natural 
volume elements $\om_0 $(respectively   $\om_1$) on $H^0( \Si_\be)$
(respectively  $H^1 (\Si_\be)$) with respect to which the torsion of 
the exact sequence (\ref{mayseq2}) is $\det (1 - \ad(g) )$, 
the determinant of the restriction to $ Z(g)^\perp \subset 
{\rm Lie}(S).$ 
Thus the contribution to the torsion from this part of 
the sequence is 
$$\frac{\om_0 \otimes \om_1 }{\Bigl \vert\det( \ad(g)- 1 ) \Bigr \vert }. $$
However,
the natural volume element $d \vol$ on
 $Sg/S$ at the point $g$ is $\det (1 - \ad(g))^{-1} \, \om_1$,
 since $1 - \ad(g) $
is the differential of the conjugation map. 
$\phi: S \to Sg$ given by $\phi(h) = hgh^{-1}. $ 
This map 
is defined because $g \in N(S).$ 
Thus, assuming the natural volume element $\om_0$ on $H^0(\Si_\be)$, 
the contribution to the torsion of $\Si_\be $ from 
(\ref{mayseq2}) is simply $d \vol \in  \lamax H^1 (\Si_\be).$ 
The part of the 
exact sequence from $H^2 \Si_\be$ to 
$H^3 \Si_\be$ gives the same contribution as 
(\ref{mayseq2}), by Poincar\'e duality.  

Thus the integral over $\calm_\rho$ of the square root of 
the torsion is 
$$ \int_{g \in Sg_0/S} \, \frac{d\vol(g)}{ \left \vert 
\det (\ad(g) \tlbe^\ast - 1 ) \right \vert^{1/2}  }. 
\bla \square $$

\section{Regularization of the eta invariant} \label{regul}

We recall the index theorem for manifolds with boundary,
whose use is necessary in order to replace the eta 
invariants which appear as phases of determinants in 
the leading order approximation to the path integral.
The general procedure is outlined in the Introduction 
(\S  \ref{introfreg}). The regularization procedure
was introduced by Witten \cite{W:J}, who 
applied it to the Chern-Simons functional. The formula 
obtained in that case is as follows: 
\beq \label{sp} 
Z(M,k) \sim 
\frac{1}{2} e^{-3 \pi i (1 + b^1(M))/4}
\:  \sum_A \tau(A)^{1/2} \, e^{-2 \pi i(I_A/4 + (\dim H^0+ \dim H^1)/8) }
\: \times \eeq
$$ \times \; e^{2 \pi i (k+h) \cs(A)} 
\, (k+h)^{(\dim H^1 -\dim H^0)/2}.  
$$
Here, we sum over the gauge equivalence 
classes of flat connections $A$. We denote the 
first Betti number by $b^1(M)$. 
To make sense of the torsion factor in (\ref{sp}), one
must choose an element of $\lamax H^0 (M;d_A)$ (which 
is done in \cite{FG}, for instance, in computing the 
torsion of lens spaces).

Here we treat the theory associated to the symplectic
action functional. 
In the SQM case, these eta invariants depend on  a choice
of complex structure at all fixed points of $f$. 
We discussed in \S \ref{s:eta} the extent to which a canonical 
complex structure could be associated to $f$. If 
$f $ preserves a compatible complex structure globally
on the symplectic manifold $N$, it is very natural to 
choose that complex structure for our calculations,
and indeed we showed in  \S \ref{hollef} that this 
choice leads to the quantity described by the 
holomorphic Lefschetz fixed point formula. For a  general
$f$,
however, it is more appealing  to seek a procedure that 
does away with the complex structure dependence altogether.

The basic ingredient in the regularization procedure
is the index theorem for manifolds with boundary:
\begin{theorem}{\rm (Atiyah - 	Patodi - Singer)}
Let $\cald$ be an elliptic operator on a manifold $X$ with 
boundary $Y$, such that in a collar $Y\times I$ near the 
boundary, $\cald$ has the form
\beq \label{formop}
\cald = \sigma(\pr/\pr t + D) \eeq
with $D$ an elliptic operator on $Y$. Then for global 
boundary conditions, 
$${\rm Index} \, \cald = \int_X \al - \frac{h + \eta(D)}{2}, $$
where $h = \dim \Ker D$ and $\eta$ is the eta invariant. 
 The differential form $\al$  on $X$ 
is the form associated to $\cald$ via the index 
theorem for closed manifolds, obtained from the asymptotic expansion of 
the heat kernel. \qqed
\end{theorem}
In particular, one  obtains a formula for 
the spectral flow of a family of operators $D_\tau$ on 
a manifold $Y$, interpreting this as the index of 
$\cald = \ddtau + D_\tau$ on $Y \times I$.
\begin{rem} The spectral flow formula above is 
{\it not} additive, if $A_0$ or $A_1$ have 
kernels: indeed, one  finds if $\cald$ is as in the previous Remark,
$${\rm Index} \, \cald[0,1] + {\rm Index  } \, \cald[1,2] = 
{\rm Index} \, \cald [0,2] - h (D_1), $$
where we have introduced $h(D) = \dim \Ker D$.
\end{rem}

\subsection{Symplectic quantum mechanics case}
Consider a family of almost complex structures $J_t$ on $TN \otimes \CC$
such that $J_{t+ 1} = f^\ast \, J_t,$ i.e.,
$$J_{t + 1} \, \circ \, df = df \, \circ J_t. $$
We obtain a corresponding family of Levi-Civita 
connections on $N$; call these $\nbt.$
Denote by $\Pi_t = \frac{1}{2}(1 - i J_t)$ the 
projection in $TN \otimes \CC$ onto the 
holomorphic tangent space $T'_t N$. Its conjugate 
$\bar{\Pi}_t $ is the projection onto 
${T''}_t N$. We define   $\bunt \eqdef \pi^\ast TN \to I \times N$, 
and denote by 
$\buntpr$ the bundle over $I \times N$ formed by the $T'_t N$. 
Because of the periodicity condition, $\buntpr$ may be
regarded as a bundle over the mapping torus $N_f$. 
A connection on $\bunt \otimes \CC$ preserving $\buntpr$ is 
given by 
\beq \label{nabdef}
\nb = \Pi_t \left (dt \ddt + \nbt\right ) \Pi_t + 
\barpit \left (dt \ddt + \nbt\right ) \barpit \eeq
$$ = \nbt + dt \left (\ddt + \Pi_t \,\tder({\Pi}_t) + 
\barpit \,\tder({\barpi}_t) \right ) $$
$$ = \nbt + dt \left (\ddt - \frac{J_t \, \dot{J_t} }{2} \right ). $$
$\nb$ commutes with  $J_t $ by construction. One easily verifies 
also that 
\begin{prop}
$\nb$ is unitary. \qqed  
\end{prop}

Consider a fixed point $x$ of $f$, which defines a 
loop $\ga_x$ in $N_f$. We  wish to study 
 $\eta(D_x)$, where $D_x$ is the operator on the bundle ${\ga_x}^\ast 
(\bunt \otimes~\CC)$ over $S^1$ given by 
\beq \label{jtddt}
D_x = J_t \, \tder. \eeq
Given two such fixed points $x_\pm$, we choose a map 
$$\tlu = \tlu(t, \tau) \; = 
\: \bigl (t, u(t,\tau)\bigr )\; : \; 
S^1 \times [-1,1] \to N_f := I \times N/\sim_f $$
 such that 
 $ u(t,\tau) = x_\pm$ in a neighbourhood of 
$\tau = \pm 1$.
In other words, $\tlu$ is a path of sections of $N_f $ 
interpolating between the constant loop at $x_+ $ and that 
at $x_-$, and $\tlu$ is 
constant in neighbourhoods of $\tau = \pm 1$. 
The  strip $\tlu$ is as in (\ref{tiludef}).

We can use $\tlu$ to pull back the bundle $\bunt \otimes \CC$
and the connection $\nb$ to $S^1 \times [-1,1]. $
Further, in neighbourhoods of $\tau = \pm 1$, the connection in 
the directions $\ddtau$ and $\ddt$ is given by 
\beq \label{utau}
(\tlu^\ast \nb)_{\tau} = \frac{\pr}{\pr \tau} \eeq
\beq \label{utt}
(\tlu^\ast \nb)_{t} = \ddt  + \nb_{\tlu_\ast \ddt} \eeq
$$= \ddt - \frac{J_t \dot{J_t} }{2}. $$

We use the standard holomorphic structure on $S^1 \times [-1,1]$, and 
define a holomorphic structure on the bundle $\tlu^\ast \buntpr  $ 
over $S^1 \times [-1,1]$ 
by defining $\dbar = (\tlu^\ast \nb)^{0,1}$. Similarly one may 
define the conjugate operator $\vardelta $ on 
$ \tlu^\ast \buntdopr $. We consider the operator $\tilde{\cald} $ 
on $\tlu^\ast \bunt \otimes \CC$ defined by 
\beq \label{curlyddef}a
\tilde{\cald} = (\tlu^\ast \nb)_\tau + J_t (\tlu^\ast \nb)_t \eeq
The condition that $u$ be constant near $\tau = \pm 1 $
was imposed to obtain (\ref{utau}), which means 
the index theorem applies to $\tilde{\cald}$ since it has the correct
form (\ref{formop}) near the boundary.  
Now $\tilde{\cald}$ restricts to $\dbar$ on $\tlu^\ast \buntpr$, 
and to $\partial$ on $\buntdopr$. 
We thus have
\beq \label{curlydind}
{\rm Index} \, \tilde{\cald} = 2 \: {\rm Index} \, \dbar\vert_{\buntpr}
\eeq
$$ = 2 \:\: \frac{i}{2 \pi} \int_u \,  \Tr F_\nb \: + 
\frac{\eta(\tld_{x_+} ) - \eta(\tld_{x_-}) } {2}
- \frac{h(x_+) + h(x_-) }{2}, $$
where here the eta invariants refer to the operator 
$\tilde{D} $ on ${\ga_x}^\ast (\bunt \otimes \CC) $ given by   
\beq \label{tildedef}
\tilde{D}_x = J_t (\tlu^\ast \, \nb)_\ddt = J_t \ddt + 
\frac{\dot{J_t} } {2}, \eeq
and we have used the notation $h(x) = \dim \, \Ker \, 
\tilde{D}_x .$
The signs of the eta invariants are because, with 
the orientation we have chosen, the boundary of $u$ is 
$\ga_{x_-} - \ga_{x_+}.$: see Lemma \ref{agre}. 

We instead wanted  the eta invariant of the operator 
\beq \label{etaorig} D_x = J_t \ddt. \eeq 
However,
$ \dot{J_t}$ is a self-adjoint bundle endomorphism of $\bunt  \otimes 
\CC$ which interchanges $ \buntpr $ and $\buntdopr$. 
Thus we may apply the following theorem to relate $\eta (D_x)  $
to $ \eta(\tilde{D}_x ) : $

\begin{theorem}{(\rm \cite{Cheeger}, (A 2.18) )} \label{cheeger}

Let $\xi_0$, $\xi_1$ be Hermitian vector bundles with Hermitian 
connections, and $L_0$, $L_1$ operators of Dirac type with 
coefficients in $\xi_0, $ $\xi_1$. Let $Q$ be 
a self-adjoint bundle endomorphism of $\xi_0 \oplus \xi_1$ 
which interchanges $\xi_0$ and $\xi_1$. For 
$\eps \in I$, define $P_\eps = 
(L_0 \oplus L_1) + \eps Q $ , an elliptic operator on $\xi_0 
\oplus \xi_1$. 
If the path $P_\eps$ consists of invertible operators, 
then $$ \eta(P_0) = \eta(P_1)  \bla \in \RR. $$
More generally, define 
$$h (P) = \dim \ker (P); $$
then 
$$\frac{\eta (P_0)+ h(P_0)}{2} =\frac{\eta (P_1)+ h(P_1)}{2} 
 \onebl \pmod{\ZZ}. \bla \square$$
\end{theorem}

\begin{rem}
Consideration of the example in \S  \ref{sltwor} shows 
that we cannot do better than this: there are choices of operators 
for which the linear path $P_\eps$ will have spectral flow 1, 
so that $\eta$ will jump by 2. For instance, take 
$$L = J (\tder + E_+), \onebl Q = JE_-$$ 
where 
$$ J = \matr{0}{-1}{1}{0}, \onebl 
 E_- = \sigone \matr{0}{1}{1}{0} 
+ \sigthr\matr{1}{0}{0}{-1}, \onebl 
E_+ =  \sigtwo \matr{0}{-1}{1}{0}. $$
In the notation of \S  \ref{sltwor}, the path $P_\eps = 
L + \eps Q$ gives  the path  of eigenvalues parametrized by $\kappa$, 
$ 0 \le \kappa \le \kappa_0$. If $ 0 < \sigtwo \ll 1, $
then the spectral flow along this path is $-1$ 
(caused by the point 
where $\la_0^- = \sigtwo - \kappa $ crosses the origin, 
see (\ref{lapm})). 
Thus  the eta invariant jumps by $2$ along this path. 
 $\square $ \end{rem}

However, Theorem \ref{cheeger} also shows that the change in the eta 
invariant is {\it only} due to the zero eigenvalues: 
\begin{prop} \label{fixdiff}
Assume $D, \tilde{D}$ have no kernel. Then the spectral flow
from $D$ to $\tilde{D}$ along a linear  path is 
$$\SF(D, \tld) = \frac{\eta(\tld) - \eta(D)}{2}. $$
More generally the index of an operator 
$\ddtau + D_\tau $ with appropriate global 
boundary conditions, where $D_\tau$ 
 interpolates linearly from $D$ 
to $\tld$, is:
\beq \label{sfd}
\SF(D, \tld) = \frac{\eta(\tld) - \eta(D) }{2} - 
\frac{h(D) + h(\tld) } {2}. \bla \square \eeq
\end{prop}

The eta invariant of $D_x$  (\ref{etaorig}) should 
be expressed in terms of the index of the operator $\cald$ 
on $\tlu^\ast (TN \times \CC)$ given as follows. We introduce
a connection $\nb' $ on $\tilde{TN} \otimes \CC$ 
which differs from $\nb $ in (\ref{nabdef}) 
by the term $ - dt \, J_t \dot{J}_t /2 $ in the 
$dt $ direction: 
\beq \label{diffs}
\nb' = \nb^{(t)} + dt \, \ddt. \eeq
Then we define 
\beq \label{calddfn}
\cald = (\tlu^\ast \nb')_\tau + J_t (\tlu^\ast \nb')_t. \eeq
Now the operators  $\cald$ and $\tilde{\cald} $ on 
$S^1 \times [-1,1] $ are homotopic (since they differ 
by a zeroth order term), but by a homotopy 
which moves their boundary values. 
We must thus 
consider the spectral flow corresponding 
to this homotopy restricting to the boundaries. 
Index $\cald(x_-, x_+) $ is the usual
index associated to the symplectic action 
functional, which is given by the difference 
of Maslov indices $\mu(x_-) - \mu(x_+)$. 

We decompose 
$${\rm Index} \, \tilde{\cald} 
 = \SF (\tld_0, \tld_1)  $$
$$ = \SF (\tld_0, D_0) + \SF(D_0, D_1) + 
\SF(D_1, \tld_1) + h (D_0) + h(D_1), $$
$$ = \frac{\eta(D_0) - \eta(\tld_0) }{2} 
 -   \frac{ h(D_0) + h(\tld_0) }{2} 
+ \SF (D_0, D_1) +  $$
$$ + \;  
 \frac{\eta(\tld_1) - \eta(D_1) }{2} 
 -   \frac{ h(\tld_1) + h(D_1) }{2} 
+ h(D_0) + h(D_1) \onebl \onebl \mbox{using (\ref{sfd}) }. $$
We equate this with the expression from (\ref{curlydind}):
\beq \SF (\tld_0,\tld_1) = \:  
{\rm Index} \: \tilde{\cald} = 2 \frac{i}{2 \pi} \,\int_u \, \Tr F_\nb 
+  \frac{ \eta (\tld_1) - \eta(\tld_0) }{2}
 - \frac{ h(\tld_1) + h(\tld_0) }{2} . \eeq
This shows, cancelling the extra terms, that
\beq {\rm Index}\: \cald = \: \SF (D_0, D_1) = 2 \, \frac{i}{2 \pi} \, 
\int \Tr F_\nb + \frac{\eta(D_1) - 
\eta(D_0) }{2} - \frac{h(D_1) + h(D_0) }{2}. \eeq
In other words, using  (\ref{sfd}), we can work backwards 
from the index formula for $\tilde{\cald}$ to the 
index formula for $\cald$. 

Our overall conclusion is
\begin{theorem} \label{compdepet}
$$- \: \frac{\eta(D_{ x_+}) - \eta(D_{ x_-})}{2} 
= 2 \:\frac{i}{2 \pi} 
\int_\tlu \Tr F_\nb  - 
 \, \SF (D_{x_-}\, ,\, D_{x_+} ) 
  - \frac{h(x_+) + h(x_-)}{2},$$
where $\tlu $ is a strip in $N_f$ interpolating between $x_+ $
and $x_-$. The curvature  $\frac{i}{2 \pi} \Tr F_\nb$
of the connection defined by (\ref{nabdef}) represents 
 $- c_1\tilde{\calk} $, where $\tilde{\calk} $ is the 
canonical bundle $ (\lamax \, \buntpr)^\ast $ over $N_f$ defined 
by the family of complex structures $J_t$. 
\qqed
\end{theorem}

\subsection{Complex structure dependence}
Notice from Theorem \ref{compdepet} 
that the difference of eta invariants still 
depends on a choice of path of complex structures on
the tangent space  
at each fixed point. 
This is in contrast to 
the situation encountered in the Chern-Simons theory, 
where {\it differences} of eta invariants are shown by 
the index theorem to be independent of a choice of 
metric. We can,
however, make sense of this situation as follows. 
The regularization procedure
we have developed  involves adding \lq\lq counter\-terms''  
of a form very similar to the \lq\lq gravitational 
Chern-Simons'' counterterm used by Witten \cite{W:J}
in regularizing the eta invariant in Chern-Simons 
theory.

We assume that 
\begin{itemize}
\item $N$ is simply connected.

\item The 
first Chern class of the canonical 
bundle $c_1 (\calk)$ $\in H^2 (N,\ZZ)$ may be 
represented by a 2-form 
$\al$ on $N$ such that $f^\ast \al = \al$.
(For example, $ \al$ might be a multiple of the 
symplectic form.)
\end{itemize}

As in \S \ref{chap1}, we may choose a line bundle with connection 
$(\lineb, \theta)$ on $N$, so that the connection $\theta$ has 
curvature specified by $ \frac{i}{2 \pi} d\theta = \al$. 

As in \S \ref{chap1},
there is then a lift of $f$ to $\tf: \lineb \to \lineb$ preserving
$\theta$: we may thus construct a bundle 
$(\tilde{\lineb}, \tilde{\theta})$
over $N_f$.
{\it As a cohomology class}, $ c_1 \tilde{\calk} $
is represented by the form $\al$ on $N_f$; thus there is 
a bundle isomorphism 
\beq \Psi: \tilde{\calk} \to \tlineb. \eeq

We thus have 
\beq \label{curvdif}
- \frac{i}{2 \pi} \Tr (F_\nb) - \al = 
\frac{i}{2 \pi} \, d ( -\Tr \nb - \Psi^\ast \, \tilde{\theta} ), \eeq
so we may replace 
$\frac{i}{2 \pi} \int_{\tilde{u}} \Tr \, F_\nb $ 
in Theorem \ref{compdepet} by 
$- \int_u \al$ by adding the \lq\lq counterterm'' 
$$\frac{i}{2 \pi} \int_{{u} }  d (\Tr \nb +
\Psi^\ast \tilde{\theta})   
 = \frac{i}{2 \pi} \left \{- \int_{\ga_{x_+} }(\Tr \nb + 
\Psi^\ast \tilde{\theta} )  + 
\int_{\ga_{x_-} }(\Tr \nb + 
\Psi^\ast \tilde{\theta} )  \right \}. $$
This means we replace each eta invariant 
$\eta(x) $ by $$ \eta(x)  -4 \,  \frac{i}{2 \pi} \int_{\ga_x} 
(\Tr \nb 
+~\Psi^\ast \tilde{\theta} ).   $$

Under this procedure, the difference 
$\eta(x_+) - \eta(x_-)$ gets regularized to 
4$\int_u \al$, where $u$ is a strip  in $N$ joining $x_-$ and 
$x_+$. The result is that instead of having 
the integral of $k$ times the symplectic form 
$\om$ in the path integral, we now have 
$ k \om + \al/2. $ This reflects the 
result in geometric quantization that one 
should alter the prequantum line bundle 
$\call$ by  tensoring with the square root
of the canonical bundle.  As we shall see 
below, in the case when $N = \calm(\Si)$ 
this yields the shift $k \to k+h$. 

There is an integer ambiguity in the data we have  
specified so far: the possible homotopy 
classes  of bundle isomorphisms $\Psi$ correspond to 
$[N_f, S^1 ] \cong \ZZ$. In fact, choosing a 
homotopy class $\Psi$ corresponds to choosing 
a homotopy class of trivializations of $\tilde{\calk} $
over 
the loop $\ga_{x_0}$ in $N_f$
for  one particular fixed point $x_0$
of $f$. We 
may see this as follows. Our lift $\tilde{f}$ was 
unique only up to a $U(1)$ factor. We may fix this
factor by requiring that $\tilde{f}$ is the 
identity over our reference 
fixed point $x_0$; then our bundle $\tlineb$ has a 
{\it canonical} trivialization over $\ga_{x_0}$. Given 
a choice of  $\Psi$, this pulls back to give a trivialization
of $\tilde{\calk}$ over $\ga_{x_0}$.

Using the gauge provided by the canonical trivialization 
over $x_0$, the integral $\int_{\ga_{x_0} } \, \tilde{\theta} $ 
is 0. Thus the regularization shifts the eta invariant 
$\eta(x_0)$ to 
$\eta(x_0) - 4 \, \frac{i}{2 \pi} \int_{\ga_{x_0} } 
\left ( \Tr  \, \nb + \nb_0 \right ), $
where $\nb_0$ is the product connection 
on $\tilde{\calk} \vert_{\ga_{x_0} } $
obtained by pulling back the canonical trivialization 
of $\tlineb $ over $\ga_{x_0}$. 
In other words, our \lq\lq counterterm'' is the transgression 
of $2 c_1(\tilde{\calk})$ in the connection $\nb$ using the pullback 
of the canonical trivialization, 
which is the precise counterpart of the gravitational 
Chern-Simons term in Witten's treatment. 

\begin{rem}
{\it  Overall phase factor:}
The number 
\beq \label{regfac} \frac{\eta(x_0) + h(x_0)}{2} - 2 \,
\frac{i}{2 \pi} \int \, \Bigl (\Tr (\nb) + \nb_0\Bigr  ) \eeq
 is an  integer, since one can use the index 
theorem and Proposition\ \ref{fixdiff} to reexpress it as 
$$\: {\rm Index } \: \tilde{\cald}[X] - 2 c_1 (X, \pi)
- \SF(D, \tilde{D}) \, - h(\tilde{D} ). $$
 Here ${\rm Index} \,\tilde{\cald} [X]$ is the index of 
the operator $\tilde{\cald} $
on the extension of the bundle $\ga_{x_0}^\ast \, 
(\buntpr \oplus \buntdopr)$  
over a surface $X$,
and $c_1(X, \pi) $ is the 
relative first  Chern class of the extended bundle 
$\ga_{x_0}^\ast \, 
\buntpr $
using a trivialization 
$\pi$ of $\tilde{\calk}$  on the boundary.
Thus, after regularization, the 
factor 
$$  \exp \, \frac{i \pi}{2} \left \{ \frac{\eta(x_0) + h(x_0)} 
{2} - 2 \frac{i}{2 \pi} \int \left ( \Tr \nb + \nb_0 
\right )  \right \} \: e^{- \frac{i \pi}{4} h(x_0)} $$
is simply a power of $i$   times $e^{- \frac{i \pi}{4} h(x_0)}$. 
In fact, recall that we had assumed our operators $D_{x_0} $
had no kernel, so  the regularized 
value of $e^{\frac{i \pi}{4} \eta (x_0) } $  is simply 
a power of $i$: we denote it $i^\mu$.  
If we change 
the trivialization $\pi$ by one unit, 
(\ref{regfac})  changes by $2$, so $i^\mu$ 
changes to $- i^\mu$.  \end{rem}

The final regularized formula for the SQM partition 
function of the manifold $N$ is thus 
\beq \label{finregsqm} 
Z (N,k) =i^\mu \,  
 e^{i k S(x_0) } \sum_{x \in N, f(x) = x} \eeq
$$  \
\exp \left \{ \; 2 \pi i \int_{u(x_0, x)} \left [k \frac{\om}{2 \pi} 
+ \frac{\al}{2} \right ] + \frac{i \pi}{2} 
{\rm Ind } \cald(x_0\, , \, x) 
\right \}  \; \; \frac{1} {\vert \det D_x 
\vert^{1/2} }.
$$

\subsection{The moduli space case}
A particular case occurs when $N$ is the moduli space 
$\calm(\Si) $ of  flat
connections   on a bundle $P_\Si $ over a surface $\Si$. 
Let $\call $ denote the determinant 
line bundle of the $\dbar $ operator on 
$\ad (P)\otimes \CC \to \Si $.
Then as a bundle over $\calm \times \calt$ (where 
$\calt$ is Teichm\"uller space), the canonical bundle 
  $\calk$ of $\calm$ may be identified with 
$\call$, provided $\dbar $ generically has 
no kernel  (see \S  4 of \cite{ADW}). We have 
$$\al =  2 h \, \frac{\om}{2 \pi}, $$ 
the curvature of the Quillen connection, where 
$\om$ is the basic symplectic form. 
The discrepancy between the Chern form 
$- \frac{i}{2 \pi}\Tr \,F_\nb$ of 
the K\"ahler 
connection and the Chern form  $\al$ of the Quillen connection is 
$$- \frac{i}{2 \pi}\Tr F_\nb - \al  
= - \frac{i}{2 \pi}\dbar \vardelta \, \log H, $$ 
where $H$ is the determinant of the $\brpr$-Laplacian.
So we 
replace $c_1 (\calk)$ by $\al$ 
and then use this to define the difference 
between eta invariants. The effect is that the coefficient 
of $\int_u \om$ multiplying 
the action (\ref{s1def}) is shifted from $k$ to 
$k + h$. 

\begin{rem} 
This argument is not valid if $\Si$ is a torus, since then the 
$\brpr$ operator generically has a kernel. However the shift 
$k \to k+h$ may be obtained by a special argument in that 
case (see \S  2 in \cite{sqpaper2}).  \qqed
\end{rem}

\subsection{Regularization: The Chern-Simons case} \label{frameta} 
For comparison, we provide here an exposition of the final step 
in the regularization procedure for the Chern-Simons theory
for a 3-manifold $M$,
following Witten \cite{W:J}. In the Chern-Simons theory, 
the relevant $\eta$ invariants are those of the Atiyah-Patodi-Singer
operator 
\beq \label{apsop}
D_A = (-1)^p (\ast d_A + d_A \ast )   \bla \mbox{ on $\Om^{2p+1}
(M, \ad (P))$.} \eeq
We denote the product connection by $A_0$. 
Here the difference $\eta (A_\al) - \eta(A_0)$ of 
two eta invariants is independent of the metric, and one 
uses the index theorem for $M \times I$ to show that it
is proportional to $\cs(A_\al)$: this leads to 
formula (\ref{sp}), and in particular to the shift of $k \to 
k+ h$ in the coefficient of $\cs (A_\al)$. 
The metric dependence comes  in the eta invariant 
$\eta(D_{A_0} )$ of the APS operator  
corresponding to the product connection.
 The 
eta invariant $\eta(A_0)$ 
is equal to $(\dim G) \eta_0$, where $\eta_0$ is the 
eta invariant of the untwisted operator $(-1)^p(\ast d + d \ast)$. 

To remove the metric
dependence of this quantity,
we choose a {\it 2-framing} $\xi$ (a trivialization 
of $2 TM$). The possible 2-framings correspond 
to $\ZZ$. The signature of a closed 4-manifold $Z$ is
$$\sign Z = \frac{1}{6} p_1 (2 TZ).$$
If $Z$ bounds a 3-manifold $M$ with a 2-framing $\xi$, we 
can define the relative Pontrjagin  class
$p_1(2TZ, \xi) \in H^4(Z, M)$. The {\it signature defect}
$\delta(M, \xi)$ of $M$ with the 2-framing $\xi$ is then 
defined by 
\begin{equation} \label{signdef}
\delta (M,\xi) = \sign (Z) - \frac{1}{6}p_1(2TZ,\xi)[Z].  
\end{equation}

By the signature theorem (a consequence of the index theorem)
we also have
\beq \label{signeta}
\sign Z = \frac{1}{6}\int_Z p_1 (2 TZ)(g) + \eta_0(g), \eeq
where both of the terms on the right hand side depend on a metric
$g$ on $M$ (the Pontrjagin form is to be evaluated in 
the curvature of the Levi-Civita connection 
on $Z$). (The  eta invariant appears with a plus sign
rather than the usual minus sign because our operator is 
$- \half$ the signature operator: see 
\cite{FG}.)
One may actually express
$$p_1(2TZ, \xi) = \int_Z p_1(F_\nb), $$
using the curvature $F_\nb$ of a connection $\nb$ on $2TZ$
which reduces on $M$ to the product connection specified
by $\xi$. Equating (\ref{signdef}) and 
(\ref{signeta}) gives 
\beq \label{defeta}
\delta = \eta_0(g) + \frac{1}{6} \int_{M \times I} p_1(F_\nb), \eeq
where now $\nb$ is a connection interpolating from the Levi-Civita 
connection of the metric $g$ to  the 
product connection defined by the 2-framing $\xi$.

So one makes sense of the factor $e^{i \pi \eta_0(\dim G)/4}$ 
by replacing it by $e^{i \pi \delta (\dim G)/4}$, which is independent
of the metric but depends on the framing $\xi$. A change in 
the framing by one unit (corresponding to a 
change in $p_1(2TZ) $ by 2) multiplies this factor by  
$e^{2i \pi \dim G/24}$. Atiyah \cite{A1} has shown 
that there is a {\it canonical} 2-framing, for which 
the signature defect $\delta$ vanishes.

\subsection{Spectral flow for SQM and Chern-Simons} \label{specflow}

The paper \cite{DS} has studied the spectral 
flow of the one-parameter family of operators 
corresponding to the gradient of the symplectic
action functional on the moduli space of flat 
connections on a surface $\Si$. One of its main 
results is that this spectral flow is equal to the 
spectral flow of the family of APS operators 
corresponding to a path of connections 
$A(\tau)$
on the corresponding
mapping torus $\Si_\be$, whose endpoints are 
flat connections $A_-$ and $A_+$ on $\Si_\be$ corresponding 
to fixed points of the action of $f_\be$ on 
$\calm(\Si)$.  Note, however, that \cite{DS} 
restricts to $SO(3)$ bundles with $w_2 \ne 0$, in order to 
obtain nonsingular moduli spaces. 
\cite{DS} also assumed that  neither 
family of operators had kernels on the 
boundary. 

In the SQM partition function (Theorem \ref{finregsqm}) we 
obtain a term 
from the spectral flow of a certain family of 
operators, similar to that in (\ref{sp}). The 
spectral flow from SQM and that 
from the Chern-Simons functional 
 enter the stationary phase approximation 
expressions with different signs, but this is simply 
the result of a discrepancy in orientation conventions.
To complete the identification of  the SQM partition 
function and the CS partition function, it would 
be 
necessary to compare the spectral flows of the APS 
operator and the operator from the gradient flow 
of the symplectic action functional, even when
these operators have kernels on the boundary.
Our operators $D_x$ are assumed to have no 
kernel, so there is no term in  the SQM stationary
phase expansion corresponding to the dimension of the  kernel
of an operator, 
in contrast to the Chern-Simons 
case (\ref{sp}).
 \qqed

\newcommand{\auth}{\sc}

\end{document}